# JavaScript Dead Code Identification, Elimination, and Empirical Assessment

Ivano Malavolta, Kishan Nirghin, Gian Luca Scoccia, Simone Romano, Salvatore Lombardi,
Giuseppe Scanniello, Patricia Lago

**Abstract**—Web apps are built by using a combination of HTML, CSS, and JavaScript. While building modern web apps, it is common practice to make use of third-party libraries and frameworks, as to improve developers' productivity and code quality. Alongside these benefits, the adoption of such libraries results in the introduction of *JavaScript dead code*, i.e., code implementing unused functionalities. The costs for downloading and parsing dead code can negatively contribute to the loading time and resource usage of web apps. The goal of our study is two-fold. First, we present *Lacuna*, an approach for automatically detecting and eliminating JavaScript dead code from web apps. The proposed approach supports both static and dynamic analyses, it is extensible and can be applied to any JavaScript code base, without imposing constraints on the coding style or on the use of specific JavaScript constructs. Secondly, by leveraging Lacuna we conduct an experiment to empirically evaluate the run-time overhead of JavaScript dead code in terms of energy consumption, performance, network usage, and resource usage in the context of mobile web apps. We applied Lacuna four times on 30 mobile web apps independently developed by third-party developers, each time eliminating dead code according to a different optimization level provided by Lacuna. Afterward, each different version of the web app is executed on an Android device, while collecting measures to assess the potential run-time overhead caused by dead code. Experimental results, among others, highlight that the removal of JavaScript dead code has a positive impact on the loading time of mobile web apps, while significantly reducing the number of bytes transferred over the network.

**Index Terms**—Dead code, JavaScript.

## 1 INTRODUCTION

Web apps are built by using a combination of HTML, CSS, and JavaScript. To increase developers' productivity via code reuse, we have been witnessing a proliferation of third-party libraries and frameworks, ranging from Model-View-Controller (MVC) frameworks, efficient DOM manipulators, User-Interface (UI) kits, etc. [1]. This phenomenon is happening not only for browser-based web apps, but even for mobile [2] and desktop software [3]. In addition to the speed-up of the development, the use of these libraries and frameworks—which are usually well tested and maintained—positively affects the quality of the implemented web-based solutions (or also *web apps* from here onwards). Unfortunately, this comes at the price of an increase in their execution time and higher usage of resources. For example, given a web app, the used JavaScript framework could include unused functionalities that are never executed. In such a context, the code implementing unused functionalities is known as *dead code* [4]. Besides the obvious cost of increased file size and network transfer time, there is an additional hidden cost to dead code: despite JavaScript dead code never being executed at run-time, it is still downloaded and parsed by the JavaScript engine. This overhead can take a significant portion of the complete execution time of JavaScript code [5]. The costs for downloading and parsing dead code can negatively contribute to the loading time and energy consumption of web apps.

While some approaches have been developed to minimize this overhead (e.g., lazy parsing[1] and script streaming[2]), dead code identification and elimination is still an open problem in web apps [6]. As far as the identification of JavaScript dead code in web apps, the currently available solutions either: *(i)* impose a certain coding style to developers, banning certain code structures (e.g., object reflection), or *(ii)* require specific constructs of the JavaScript specification. An example of the latter is the use of modules, which allow developers to specify self-contained namespaces in JavaScript and to conditionally load them when needed. While modules are certainly useful in terms of maintainability and code reuse, most web apps today have not been built with modules in mind [1].

Researchers have investigated the presence of dead code in web apps. For example, Boomsma *et al.* [7] reported that, in a subsystem of an industrial web app written in PHP, the developers removed 30% of the subsystem's files because these files were actually dead code. Eder *et al.* [8] observed that, in an industrial web app written in .NET, 25% of methods were dead. Surprisingly, no empirical studies have been conducted to assess the effect of JavaScript dead code on Web apps at run-time. For example, so far, no

- *I. Malavolta, K. Nirghin, and P. Lago are with the Vrije Universiteit Amsterdam, The Netherlands.*
  *E-mails: i.malavolta@vu.nl, k.j.nirghin@student.vu.nl, p.lago@vu.nl*
- *G. L. Scoccia, is with the University of L'Aquila, Italy.*
  *E-mail: gianluca.scoccia@univaq.it*
- *S. Romano and G. Scanniello are with the University of Salerno, Italy*
  *E-mail: siromano@unisa.it, gscanniello@unisa.it*
- *S. Lombardi is with the University of Basilicata, Italy.*
  *E-mail: salvatore.lombardi@studenti.unibas.it*

1. https://v8.dev/blog/preparser
2. https://v8.dev/blog/v8-release-75

empirical studies have been conducted to assess the impact of downloading and parsing the JavaScript dead code of a web app. In other words, the common belief is that there is a cost to pay when JavaScript dead code is present, but there is no evidence of its extent.

The goal of this paper is two-fold. First, we present *Lacuna*, an approach for automatically eliminating JavaScript dead code from web apps (Section 2.2). Secondly, we empirically evaluate the run-time overhead of JavaScript dead code in terms of energy consumption, performance, network usage, and resource usage in the context of mobile web apps (Section 3).

**Lacuna.** At the core of Lacuna lies the construction of a call graph $G_w$ of the web app $w$ being analysed; $G_w$ is undirected and represents JavaScript functions as nodes and the caller-callee relationship between functions as edges. In this context, dead code elimination consists of the removal of all the (connected) components in $G_w$ that are isolated from the root node representing the global scope of the web app. The unique characteristic of Lacuna is its ability to *build and iteratively refine* $G_w$ by executing in sequence different program analysis techniques, each with its own potential support for specific aspects of the JavaScript language. Lacuna supports any kind of program analyses (both static and dynamic), provided that they are aimed at building a call graph of the JavaScript code being analysed. Lacuna is *extensible* and independent from the used program analysis techniques, allowing developers and researchers to build the combination of analyses that best fits their own needs. Finally, Lacuna can be applied to any JavaScript-based web app, without imposing any constraints on the developer on coding style (e.g., banning the use of reflection or objects self-inspection) or on the use of specific JavaScript features (e.g., modules). We exploit this feature of Lacuna in our experiment, where we assess the run-time overhead of JavaScript dead code on 30 independently-developed mobile web apps.

**Experiment.** The goal of our experiment is to empirically assess the overhead that JavaScript dead code has when executing mobile web apps. We scope this experiment in the context of *mobile web apps* since *(i)* web browsers are more used on mobile devices [9], *(ii)* the web browser is one of the most used apps on mobile devices [10], and *(iii)* mobile devices tend to have limited processing power, poorer network capacities, and lesser memory with respect to desktop machines [5]. In this experiment, we target 30 mobile web apps independently developed by third-party web developers. The 30 web apps are divided into two different families: 15 *in-the-lab* web apps and 15 *in-the-wild* web apps. In-the-lab subjects are randomly sampled from the TodoMVC project [11]. This project contains different implementations of the "same" Todo web app, each using a different JavaScript MV* (Model View Anything) framework (e.g., AngularJS, React, Vue.js, etc.). Since all in-the-lab subjects share the same functionalities, they might negatively influence the experiment's external validity, making our results less generalizable. In order to mitigate this potential bias, we decided to complement the 15 in-the-lab subjects with 15 additional in-the-wild subjects; those subjects are sampled from the Tranco list [12] and include well-known web apps such as amazon.com, wikipedia.com,

and youtube.com. We applied Lacuna four times on each of the 30 mobile web apps, each time eliminating dead code according to a different optimization level of Lacuna (see Section 2.2.4). Later, we executed each different version of the mobile web apps, while collecting measures where the presence of dead code might result in run-time overhead for the user experience or for the (technical, ecological) sustainability of mobile web apps. The most notable results of this experiment are:

- eliminating JavaScript dead code makes the considered mobile web apps *slightly more energy-efficient* across all Lacuna optimization levels, but this phenomenon is not statistically significant;
- considered mobile web apps load faster when dead code is eliminated, especially for the most aggressive optimization level of Lacuna (this result is statistically significant, with a small effect size), however, the measures of *first contentful paint* and *first paint* do not show any noticeable improvement across the various Lacuna optimization levels;
- the elimination of JavaScript dead code leads to noticeable (and statistically significant) differences in terms of the number of performed HTTP requests only for in-the-lab subjects;
- the number of transferred bytes (significantly) diminishes when dead code is eliminated, especially for the most aggressive optimization level of Lacuna, with small effect size for in-the-lab subjects and medium effect size for in-the-wild subjects;
- CPU and memory usage tend to be (significantly) lower when dead code is eliminated from in-the-wild subjects, but not for the in-the-lab subjects; GPU usage is (significantly) lower for in-the-lab subjects without JavaScript dead code, but not for the in-the-wild ones.

An initial version of Lacuna was presented at the 2018 IEEE International Conference on Software Analysis, Evolution and Reengineering [6]. The first new contributions of this journal version consist of an in-depth description of the new features of Lacuna, described in Section 2.2.6. The current implementation of Lacuna has been completely redone and it is publicly available on itHub [13]. Another new contribution of this paper is the empirical evaluation of Lacuna, for which we designed, conducted, and reported an experiment about the run-time overhead of JavaScript dead code in terms of energy consumption, performance (e.g., page load time), network usage, and resources usage in the context of mobile web apps. In summary, the **main contributions** of this paper are:

- the presentation of Lacuna, an extensible approach for JavaScript dead code elimination;
- the integration of five new third-party analysis techniques in Lacuna;
- a completely new and publicly available implementation of Lacuna in Node.js;
- an experiment on the run-time overhead of JavaScript dead code on 30 third-party web apps;
- a publicly available replication package [14].

The **target audience** of the research presented in this paper consists of *(i)* web developers and *(ii)* researchers. Web developers can use the current implementation of Lacuna for removing dead code from their web apps, thus making their products more lightweight in terms of, e.g., network usage, load time, or energy consumption. Researchers can use Lacuna as a means for benchmarking their analysis techniques for JavaScript dead code elimination.

**Paper Structure.** The remainder of this paper is organized as follows. Section 2.1 provides background information on dead code, while Section 2.2 presents our extended version of Lacuna. In Section 3, we introduce the empirical study on the run-time overhead of JavaScript dead code. The obtained results are reported in Section 4. A discussion of the obtained results and the threats to validity of the experiment are presented in Section 5. Finally, Section 6 presents related work and Section 7 closes the paper with final remarks.

## 2 BACKGROUND

In this section, we provide context and discuss preliminary concepts required in the remainder of our paper. We define the concept of dead code, discuss related research, provide a description of the inner workings of Lacuna, and describe the results of our internal and external evaluations of Lacuna.

### 2.1 Dead Code

Dead code is part of the so-called *code smells*, a series of indicators and characteristics in the source code of a program that can possibly indicate a deeper problem. Although dead code was not considered by Fowler in his original catalog of code smells [15], it was later introduced in their respective code-smell catalogs by Brown [16], Wake [17] and Martin [18].

The perspective of these authors towards dead code is that of *refactoring*—i.e., dead code removal makes source code easier to comprehend and maintain [19], [20]. Developers, besides being interested in dead code for refactoring reasons, can be interested in *optimization* and *energy-efficiency* reasons. In other words, developers do not remove dead code because they are interested in improving source code comprehensibility and maintainability, but because they want to make their apps faster and/or lighter (optimization reason) or less energy-consuming (energy-efficiency reason). This perspective, that is the one taken in this paper, has practical implications: if the perspective is of refactoring, the removal of dead code does not regard external dependencies (e.g., libraries or framework); if the perspective is of optimization or energy-efficiency, developers need to remove dead code from external dependencies as well. Specifically, we adopt in our paper the optimization and energy-consuming perspectives when detecting and removing JavaScript dead code. Accordingly, we are not interested in the benefits, deriving from dead code removal, in terms of source code comprehensibility and maintainability; moreover, we remove JavaScript dead code from external dependencies as well.

A survey among almost 9,300 JavaScript developers rated code splitting and dead code elimination as the highest-rated requested features [21]. However, due to the highly-dynamic and event-based nature of JavaScript, it is hard to completely and correctly analyze JavaScript source code [22]. The features of this language pose challenges for analysis tools, making call-graph construction[3] and dead-code removal especially difficult. To circumvent these challenges, currently available tools for the detection and removal of JavaScript dead code tend to prevent the use of language features (such as reflection) or require the application to meet certain characteristics. Bundlers like rollup.js and Webpack perform dead code elimination using a process known as tree-shaking [23]. This is an effective way of (partial) dead code elimination. Differently from Lacuna, tree-shaking requires the use of ECMAScript6 modules, which are not widely supported at the time of writing [24]. Moreover, it requires developers to meticulously write import and export statements, as otherwise unused functions might still be imported. The Google Closure Compiler is a tool that rewrites JavaScript code to improve download and execution speed. It analyzes the source code, removes dead code, and rewrites it to a more optimal form [25]. While the Closure Compiler is effective for dead code elimination, it requires, differently from Lacuna a specific coding style. Recently Kupoluyi *et al.* [26] propose Muzeel, a black-box approach (to identify and remove dead code functions in JavaScript libraries) that requires neither knowledge of the code nor execution traces. To identify dead code functions, Muzeel performs dynamic analysis through an user's emulation implemented in a bot (i.e., browser automated tool). One of the most remarkable differences with Muzeel is that Lacuna combines source code analysis and dynamic approaches to identify dead code functions and this allows saving computation time to their identification.

The call graph representation of JavaScript programs is the base of many static analysis tools; not only for the detection of dead code but also to detect security issues [27]. For example, Antal *et al.* [27] compare five widely adopted static tools. In addition to (Google) Closure Compiler, the authors analyze npm cg, WALA, Approximate Call Graph (ACG), and Type Analyzer for JavaScript tools (TAJS). The authors observe a variance in the results of these tools (in terms of number, precision, and type of call edges) and suggest combining their output to get a better trade-off in the construction of call graphs. Chakraborty *et al.* [28] identify different root causes of missed edges in JavaScript static call graphs and an approach to build call graph representations of JavaScript programs. The approach works by identifying the dynamic function data flows relevant to each call edge missed by the static analysis. In the implementation of Lacuna, we take advantage of the findings reported in [27], [28] by combining the results of different static and dynamic analyzers to obtain a single call graph representation of a JavaScript program (see Section 2.2.5)

As compared with past research, Lacuna is based on both static and dynamic analyses, it can be easily extended, and it can be applied to any JavaScript code base (e.g., without imposing constraints on the developers' coding style). In this paper, we also present the results of an empirical assessment of the possible run-time overhead of JavaScript dead code in terms of energy consumption, performance, network usage, and resource usage in the context of mobile web apps. The experimental subjects of our assessment were 30 third-party web apps (e.g., amazon.com, wikipedia.com, and youtube.com) that we have run on a real Android device.

---

3. A call graph contains nodes that represent functions of the program and edges between nodes if there exists at least one function call between the corresponding functions.

**Running example**. In the remainder of this section, we will adopt the sample program of Listing 1 as a running example. Three functions compose the example program: function *a* (lines 1-5 in Listing 1) is directly invoked from the global scope (line 16) and thus is reachable; function *b* (lines 7-9) is reachable as is called from function *a* after a timeout has expired (lines 2-4); function *c* (lines 11-14) is not called by any other function and thus is unreachable and represents dead code.

```
1   function a(){
2     setTimeout(function(){
3       b();
4     }, 6000);
5   }
6
7   function b(){
8     console.log('6 seconds have passed');
9   }
10
11  function c(){
12    console.log('function c has been called');
13    /* Other potentially heavy statements */
14  }
15
16  a.call();
```
Listing 1: Running example

## 2.2 Lacuna

In this section, we describe the inner workings of Lacuna. The high-level workflow of Lacuna is outlined in Figure 1.

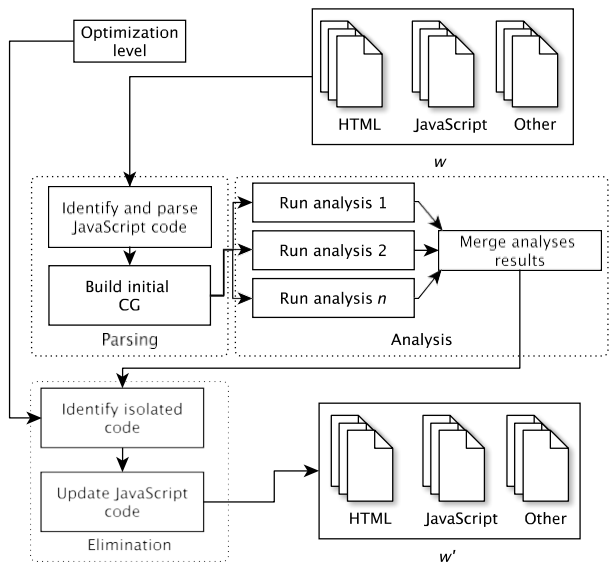

Fig. 1: Workflow of Lacuna

Lacuna takes as input *w*, the source code of the web app being analyzed, and *l*, the desired optimization level. It is important to note that these are the only input needed by Lacuna, making it applicable in the context of a wide spectrum of projects, independently of the used development process or company-specific practices. In the first phase, namely the **Parsing** phase, JavaScript code inside *w* is detected and parsed, and an initial Call Graph (CG) is built. The results of this phase are provided as input to the second phase, **Analysis**, in which multiple analysis techniques integrated in Lacuna are executed in parallel and the results of each of them are merged. Finally, the last phase, **Elimination**, is executed. In this phase, dead code is identified and the corresponding JavaScript source code is optimized according to the optimization level *l*. The final output is *w'*, an optimized version of *w* where the detected dead code is removed.

In the following of this section, we first introduce preliminary concepts and then the three phases (i.e., Parsing, Analysis, and Elimination) behind Lacuna. Finally, we provide implementation details, including used technologies.

### 2.2.1 Preliminary concepts

All the algorithms adopted in the parsing, analysis, and elimination phases operate on call graphs. A *call graph* $G = (V, E)$ is a uni-directed graph where the set of nodes $V$ represents JavaScript functions and the set of edges $E$ represents the caller-callee relationship between functions. Specifically, an edge $e_{ij}$ between the node $i$ and the node $j$ in $G$ represents the fact that the function $i$ is able to call the function $j$. In the context of JavaScript web apps, a call graph always contains one root node; such a root node corresponds to the JavaScript global scope, which is always present and executed when the web app is run in the browser.[4]

In this context, dead code elimination consists of the removal of all the (connected) components in $G$ that are isolated from the root node representing the global scope of the web app. Due to the highly-dynamic and event-based nature of JavaScript, the identification of the edges of $G$ is difficult [22], [29], [30]. As explained in Section 2.1, currently there is no technique for building correct and complete call graphs for JavaScript without imposing any constraints to developers or making strong assumptions on the usage of the language, e.g., having a complete test suite or prohibiting the use of reflection. To overcome this challenge, Lacuna leverages a set of external analysis techniques, $A$. Lacuna considers the included analysis techniques as black-box components, with the only assumption being that each analysis technique $a \in A$ adheres to the interfaces defined by Lacuna, meaning that it has to take as input an initial call graph $G_0$ and the source code of *w*, and builds its own call graph $G_a$, leveraging principles and analysis techniques of choice for the identification of edges. This allows for the inclusion of analysis techniques that are either dynamic or static. In Section 2.2.5, we will show that this restriction is not limiting and several existing tools have been integrated into Lacuna with relatively low effort. Each edge in $G_a$ will be labeled with the analysis technique that identified it. Thus, in our final call graph, built from the combination of all graphs $G_a$s, each edge can have multiple labels to take into account the fact that multiple analysis techniques can identify the same function call as reachable.

4. http://www.w3schools.com/js/js_function_invocation.asp

### 2.2.2 Parsing

In the parsing phase, Lacuna performs two main procedures, **Parse** and **InitializeCG**, both described in Algorithm 1. In the first, given as input the source code of the web app being analyzed $w$, Lacuna identifies all the JavaScript code within it by considering *(i)* all the JavaScript code defined in-line in all HTML files, *(ii)* all JavaScript files referenced by the HTML code by means of the <script> tag, and *(iii)* all the JavaScript files in $w$ that are not referenced by any <script> tag (lines 1-6 in the Algorithm 1). Once all the JavaScript code related to $w$ has been identified, Lacuna parses it into an internal representation of all its statements to ease subsequent steps (lines 7-16). During this step, to enable full analysis and optimization of $w$, all the externally hosted JavaScript code will be downloaded locally (lines 9-11). With the assumption that the entirety of the program is contained in a single example.js file, this first phase is trivial for our running example of Listing 1.

---

**input** : *w,* source code of the web app to analyze
**output**: $G_0$, initial call-graph representation of $w$

1 **Function** *Parse(w)* → *S:*
2 **begin**
3     $J_{inline}$ = JavaScript code defined in-line in $w$
4     $J_{script}$ = JavaScript code in files referenced by <script> tags in $w$
5     $J_{file}$ = JavaScript code in files not referenced by any <script> tag in $w$
6     $J = J_{inline} \cup J_{script} \cup J_{file}$
7     $S = \varnothing$
8     **foreach** $j \in J$ **do**
9         **if** *j is externally hosted* **then**
10             download *j* locally
11         **end**
12         $s$ = statements in $j$
13         $S = S \cup \{s\}$
14     **end**
15     **return** $S$
16 **end**

17 **Function** *InitializeCG(S)* → $G_0$:
18 **begin**
19     $G_0 = (V_0 = \varnothing, E_0 = \varnothing)$
20     **foreach** *function declaration f in S* **do**
21         $V_0 = V_0 \cup \{f\}$
22     **end**
23     $V_0 = V_0 \cup \{global\}$
24     **return** $G_0$
25 **end**

**Algorithm 1:** Parsing Algorithm of Lacuna

---

Afterward, as part of the InitializeCG procedure, an initial call graph $G_0 = (V_0, E_0)$ is instantiated. To this end, first, all function definitions within $w$ are retrieved, including anonymous and inline functions, and a node for each identified function declaration is created in $G_0$ (lines 17-22). Additionally, a starting node representing the JavaScript global scope is included in $G_0$ to be able to consider also all those functions directly called from the global scope of the web app (line 23). The $G_0$ call graph of our running example contains five nodes, namely: the *global* node, one node for each $a$, $b$, and $c$ functions, and one node for the inline function defined in the setTimeout call. $G_0$ does not contain any edge in this phase, they will be added in the next phase.

### 2.2.3 Analysis

Lacuna's analysis algorithm is presented in Algorithm 2, once more divided in the **Analyze** and **Merge** procedures. The former takes as input the $G_0$ call graph and produces as output a set of call graphs $H$. To do so, it executes each $a \in A$ in parallel on $G_0$ and collects each resulting $G_a$ in $H$ (lines 1-10 in Algorithm 2). During its execution, each $a \in A$ performs the identification of the edges in $G_0$, leveraging its own analysis principles. For instance, TAJS relies on abstract interpretation, which approximates the execution of a program (and thus the identification of edges) by means of monotonic functions [31], while ACG employs a field-based flow analysis technique, which statically approximates the flow of data [32]. Let us refer to the running example of Listing 1, and let us assume, for the sake of a simpler explanation, that among the analysis techniques available in Lacuna (see Section 2.2.5), only the following three are executed: *static*, *dynamic*, and *native calls*. Each of the three techniques is executed independently on $G_0$, producing the set of call graphs $H = \{G_{static}, G_{dynamic}, G_{nativecalls}\}$. These techniques, and the other ones available, are explained in detail in Section 2.2.5.

After all the analysis techniques have been executed, during the *Merge* procedure Lacuna joins the call graph $G_a$ produced by each analysis technique $a$ into a final call graph $G_w$. The strategy applied in this step is the following: *(i)* each node in $G_0$ is replicated into $G_w$ (line 13), *(ii)* for each $G_a \in H$ we add all its edges into $G_w$ (lines 14-17), *(iii)* when adding an edge $e_{ij}$ produced by a technique $a$, if $e_{ij}$ is already in $G_w$, then we just add the label $a$ to $e_{ij}$ (lines 18-21). The resulting graph $G_w$ is provided as output.

Figure 2 shows the merged call graph $G_w$ for our running example after running the three analysis techniques mentioned in the previous paragraph. Here, the *dynamic* analysis identified the call from the *global* scope to the function $a$, the *native calls* analysis identified the call from $a$ to the inline function defined in lines 2-4 of Listing 1 (by considering the call to setTimeout as a direct function call), and the *static* analysis identified the call from the body of the previous function definition to $b$. No analysis technique identified any call to function $c$, so it is unreachable from *global* because it has no incoming edges at all.

### 2.2.4 Elimination

Once all analysis techniques have been executed and the complete $G_w$ is available, the elimination phase identifies all the nodes in $G_w$ representing dead code. The algorithm employed by Lacuna in this phase is presented in Algorithm 3, constituted by the **IdentifyAlive** and **RemoveDead** procedures. The IdentifyAlive procedure identifies alive nodes in $G_w$. To do so, it performs a traversal of $G_w$, starting from the root node *global*, while keeping track of $G_v$, the graph of visited nodes (lines 1-5 in Algorithm 3). Nodes visited during this traversal are knowingly alive, as *(i)* there exists a path of edges in $G_w$, representing caller-callee relationships,

```
input  : w, source code of the web app to analyze
         G₀, initial call-graph representation of w
output : Gw, complete call graph of w
```

**1 Function** *Analyze(w, G₀) → H:*
**2 begin**
**3**    $A$ = set of analysis techniques in Lacuna
**4**    $H = \varnothing$
**5**    **foreach** $a \in A$ **do**
**6**        $G_a$ = result of execution of $a$ on $\langle w, G_0 \rangle$
**7**        $H = H \cup G_a$
**8**    **end**
**9**    **return** $H$
**10 end**

**11 Function** *Merge(H, G₀) → Gw:*
**12 begin**
**13**    $G_w = (V_w = V_0, E_w = \varnothing)$
**14**    **foreach** *graph* $G_a \in H$ **do**
**15**        **foreach** *edge* $e_{ij} \in E_a$ **do**
**16**            **if** $e_{ij} \notin E_w$ **then**
**17**                $E_w = E_w \cup \{e_{ij}\}$
**18**                $label(e_{ij}) = a$
**19**            **else**
**20**                $label(e_{ij}) = label(e_{ij}) \cup \{a\}$
**21**            **end**
**22**        **end**
**23**    **end**
**24**    **return** $G_w$
**25 end**

**Algorithm 2:** Analysis algorithm of Lacuna

```
input  : w, source code of the web app to analyze
         Gw, complete call graph of w
         l, desired optimization level
output : w', optimized version of w
```

**1 Function** *IdentifyAlive(Gw) → Gv:*
**2 begin**
**3**    $G_v$ = result of $G_w$ traversal starting from *global*
**4**    **return** $G_v$
**5 end**

**6 Function** *RemoveDead(w, Gw, Gv, l) → w':*
**7 begin**
**8**    $w' = w$
**9**    **foreach** *node* $n \in (V_w - V_l)$ **do**
**10**        retrieve $f$ = function declaration of $n$
**11**        **if** $l = 0$ **then**
**12**            $w'(f) = f$, no change to $f$
**13**        **else if** $l = 1$ **then**
**14**            $w'(f) = f_{lazy}$, lazy-loading version of $f$
**15**        **else if** $l = 2$ **then**
**16**            $w'(f) = f_{empty}$, empty-body version of $f$
**17**        **else if** $l = 3$ **then**
**18**            ~~**Algorithm 3:** Elimination algorithm of Lacuna~~
**19**        **end**
**20**    **end**
**21**    **return** $w'$
**22 end**

from the *global* node to each of these nodes, and *(ii)* we consider the *global* node as always alive since in JavaScript the global scope is always present and executed. Hence, we consider dead code each node $n$ in $G_w - G_v$, a subgraph of all disconnected components unreachable from the *global* node. Nodes in this sub-graph represent JavaScript functions that are not called by any other function (or from the global scope) according to all the different analysis techniques applied in the previous phase. For our example of Listing 1, almost all nodes would be visited during the traversal of the graph $G_w$, which starts from the *global* node. The sole exception is represented by the node $c$, which does not have any incoming edge and thus it is a disconnected component unreachable from the *global* node (visible in Figure 2).

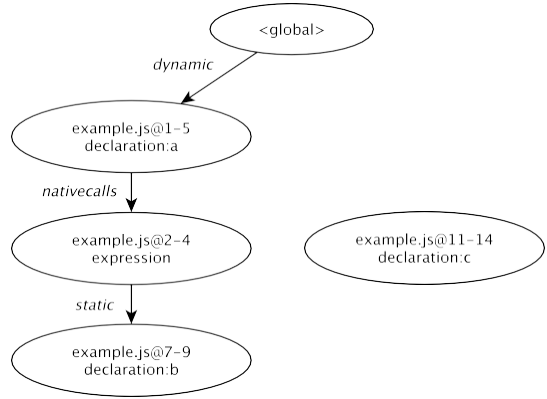

Fig. 2: Call graph of running example constructed by Lacuna

Afterward, in the RemoveDead procedure, Lacuna reconsiders the source code of the web app $w$ and performs removal of the JavaScript functions corresponding to nodes in $G_w - G_v$ (lines 12-22). This step is performed according to the user-selected optimization level, among four increasingly aggressive choices. Figure 3 provides an example of code optimized with each level. Available levels are:

- **Optimization level 0:** does not perform any optimization and leaves the source code of the web app intact. This may be useful for users to gain insights into which functions may be dead without actually removing them. Thus resulting code is the same as the one given as input (Figure 3a). Information about dead and alive functions is given in Lacuna log files.
- **Optimization level 1:** replaces the function body with a lazy-loading mechanism that will retrieve the original function body when called, following the technique suggested by Vazquez *et al.* [33]. Lazy-loading effectively is a fall-back that ensures that the application will not break due to the presence of false positives. For the lazy-loader to work, a lazy-load server containing all removed function bodies needs to be running at all times. The lazy-loading code will make an HTTP request to the server to fetch the original function body when needed. An example of performed optimization is shown in Figure 3b, where the original body of the dead function $c$ is replaced by a call to the lazy loader LazyLoad (line 20), that dynamically fetches the original function body from the lazy loading server when invoked (lines 12-16).

- **Optimization level 2:** performs a conservative optimization, removing the function body while keeping the function declaration. The rationale for this choice is that in many cases function declarations are used as expressions in JavaScript and are used in various contexts in which complete removal would lead to run-time errors in the browser. Figure 3c shows an example of code optimized with this level, where the original function body of c has been removed (line 18) but references to it have been preserved (line 22).

- **Optimization level 3:** performs an aggressive optimization, removing the presumed dead functions entirely. This elimination strategy maximizes the benefits of dead code removal. However, it also maximizes adverse effects in the case of false positives. In Figure 3d, we provide an example of code optimized at level 3, where the dead function c has been removed entirely (line 17), including references to it (line 22).

After applying optimizations, *w'*, the optimized version of the web app provided as input, is returned as output to the user. With the exception of optimization level 0, all optimizations are applied on a copy of the original web app *w* provided as input. In our example of Listing 1, the proper optimization, in accordance with the user-defined optimization level, would be applied only for the function *c*.

### 2.2.5 Implementation and used technologies

We developed Lacuna as a Node.js application. To carry out parsing of the input web app *w*, we adopt the Esprima [34] parser. Currently, our implementation comes with eight ready-to-use analyzers, which have already been integrated into Lacuna. Each is described in the following:

- **Static**: a static analyzer based on an approach utilizing point analysis [35]. It makes use of Esprima, and builds an approximate call graph, ignoring dynamic properties and context binding.

- **Dynamic:** a basic dynamic analyzer for web apps. Firstly, it instruments the web app by adding logging statements at the beginning of the body of every function definition (including anonymous and inline functions). Then, it runs the web app in a headless browser (namely, in our implementation, PhantomJS [36]), collects the logging information, and builds the call graph according to the functions executed at run-time. It does not provide any input to the web app while executing it.

- **Native calls:** an extension of the ACG analyzer, where we also consider native JavaScript functions (i.e., Array.prototype.map or Function.prototype.call) when building the call graph.

- **ACG:** our implementation of the field-based call graph construction algorithm proposed by Feldthaus *et al.* [32]. It does not consider dynamic properties, and it does not take arrow functions into account.

- **WALA [37]:** a static analysis framework for Java and JavaScript. It builds an intermediate form of the JavaScript code being analysed, then used as a basis for pointer analysis and call graph construction. We wrapped the publicly available implementation [37] in a Lacuna module.

- **TAJS:** a dataflow analysis technique for JavaScript that infers type information and call graphs [38]. TAJS performs abstract interpretation using a customization of the monotone framework [39] tailored to precisely model JavaScript-specific constructs [40]. We wrapped the Java implementation [41] in a Lacuna module.

- **npm_cg [42]:** npm cg is a tool made to produce call graphs from JavaScript source code. It comes with a series of significant limitations: only a single JavaScript file is considered at a time and only named JavaScript functions are taken into consideration (thus no arrow functions or function expressions are considered). Minor modifications were made to its implementation to integrate it in Lacuna. The resulting implementation is available in the Lacuna repository [13] along with a patch file reflecting all changes made to the original source code.

- **Closure Compiler:** the Closure Compiler [25] is a tool from Google for making JavaScript download and run faster. Instead of compiling from a source language to machine code, it compiles from JavaScript to an improved JavaScript where dead code is removed and live code is minimized. Behind the curtains, the closure compiler creates a call graph for its internal representation of the source code. By default there is no way of outputting this call graph, therefore some modifications were made to output the call graph. The resulting implementation is included in the Lacuna repository [13] along with a patch file that reflects the changes made to the original source code.

It is worth mentioning that the Static, Dynamic, WALA analyzers were already integrated into the previous version of Lacuna [6]. On the other hand, we integrated ACG, TAJS, npm_cg, and Closure Compiler into Lacuna because these analyzers were empirically assessed in the comparative study by Antal *et al.* [27], who concluded that combining more analyzers, rather than using them individually, can lead to more accurate JavaScript call graphs. Finally, we included Native calls since it is an extension of ACG.

### 2.2.6 Novel features and extensions

An initial version of Lacuna was presented at the 2018 IEEE International Conference on Software Analysis, Evolution, and Reengineering [6]. With respect to the previous paper, novel features of Lacuna presented in this journal version include:

- the new subsystem for the removal of dead code according to four different optimization levels, previously described in Section 2.2.4;

- the support for externally hosted JavaScript code. Externally hosted JavaScript files are now downloaded locally during the parsing phase, as to enable a correct and complete analysis of the application under scrutiny. This enables the analysis of web

Fig. 3: Optimization levels offered by Lacuna: (a) original example code, (b) after applying optimization level 1, (c) after applying optimization level 2 and (d) after applying optimization level 3.

apps partially hosted on a public Content Delivery Network (CDN);

- the support for JavaScript code embedded into non-JavaScript files—i.e., the HTML files of the web app under analysis are now considered during the parsing phase, to identify JavaScript code referenced or embedded by them;

- the integration of five new third-party analysis techniques into Lacuna, namely ACG [32], TAJS [38], npm cg [42], Native calls, and Closure Compiler [25];

- improvements to the JavaScript call graph representation. Specifically, in the new version of Lacuna, each call graph node (i.e., code function) is annotated with a number of supplemental information (e.g., source file name, starting code line, ending code line) to allow for easier integration of the output of different analysis techniques;

- the new version of Lacuna is made available as a stand-alone NodeJS module, which can be imported into any NodeJS project. This allows for easier integration of Lacuna into a development pipeline.

### 2.2.7 Correctness, Completeness, and Accuracy of Lacuna

We empirically assessed Lacuna in terms of correctness, completeness, and accuracy of the detected JavaScript dead functions. To do so, we replicated our previous experiment [6] on a wider dataset and by considering more instances of Lacuna—each instance either uses a single analyzer to build JavaScript call graphs or a combination of analyzers. The dataset of this experiment consists of 39 web apps developed by independent web developers in the context of the TodoMVC project—as compared to our previous experiment on Lacuna [6], we included 10 more web apps. Later, for each web app, we built the ground truth (i.e., we determined which functions are actually dead or alive), executed each instance of Lacuna on it, and then gathered the functions detected as dead by that instance. In total, we ran 127 different instances of Lacuna, each instance integrated one to seven analyzers—in our previous experiment on Lacuna [6], we ran three instances of Lacuna only: Static, Dynamic, and their combination. The analyzers we executed in this replicated experiment are those listed in Section 2.2.5 with the exception of WALA.[5] Finally, we quantified the correctness, completeness, and accuracy of the detected JavaScript dead functions by using the *precision*, *recall*, and *F-score* measures from the Information Retrieval (IR) field [43]. The results of this experiment suggest that: *(i)* combining two or more analyzers leads to improvements in terms of correctness, completeness, and accuracy of the detected JavaScript dead functions and *(ii)* the best instance of Lacuna is the one based on the joint use of Dynamic and TAJS. This instance allows achieving the highest accuracy level (average F-score = 87.9%) so well balancing correctness (average precision = 82.5%) and completeness (average recall = 97.2%). While F-score is a trade-off measure between precision and recall, the average values of precision and recall reported above can be interpreted as follows: on average, 82.5% of the functions this instance of Lacuna detects as dead are truly dead (i.e., on average, only 17.5% of the functions are wrongly detected as dead); on the other hand, this instance of Lacuna detects 97.2% of all the dead functions available in a web app (i.e., on average, it misses less than 3% of the dead functions available in a web app). It is important to note that having a precision of 82.5% might be acceptable for many projects, but it might be not enough for some other projects (e.g., those where the incorrectly removed dead code performs critical functionalities of the web app); in the latter case we suggest to the users of Lacuna to adopt optimization level 1, where the body of incorrectly-removed functions is lazily loaded and executed from a server [33]. We also suggest the users of Lacuna to experiment with other combinations of analyzers, which might lead to a precision-recall combination which is better fitting the requirements of their project and organization. For example, in our experiment the combination of the Dynamic, Closure Compiler npm cg, TAJS, and ACG lead to a higher precision (88.1%) than the one obtained via the Dynamic-TAJS combination (82.5%); however, the higher precision came with a high cost in terms of recall, which was only 54.3%, thus leading to a much lower F-score (64.8%). Thanks to the extensible architecture of Lacuna, in those cases where already-existing analyzers do not perform well, developers can still integrate in Lacuna a new analyzer with their own project- or organization-specific algorithms for building more accurate call graphs. Nevertheless, at the time of writing, as we will report in Section 2.2.8, the aggregated F-score values achieved by Lacuna with the Dynamic-TAJS combination are the highest when compared to those of other state-of-the-art approaches.

---

5. When executing the WALA analyzer on the selected web apps, it turned out to be too slow—the execution times of WALA were even greater than 20 minutes for most of the web apps. Therefore, we discarded WALA (including any combination with other analyzers) from this experiment.

TABLE 1: Average values regarding the correctness, completeness, and accuracy of the old Lacuna version—the values are those reported in the previous Lacuna paper [6]

| Variable | Static | Dynamic | Static + Dynamic |
|---|---|---|---|
| Precision | 56% | 57% | 63% |
| Recall | 49% | 77% | 40% |
| F-score | 49% | 64% | 47% |

Finally, to give an idea about the impact of the improvements to Lacuna, we summarize in Table 1, the average values of the F-score, precision, and recall measures reported in the previous Lacuna paper [6], where three instances of the old Lacuna version were studied (i.e., Static, Dynamic, and their combination). It is easy to grasp that the new Lacuna version, based on the combination of Dynamic and TAJS, lead to improvements in terms of correctness, completeness, and accuracy of the detected JavaScript dead functions. For details about the replicated experiment briefly described in this section, we redirect the interested reader to our online appendix [44].

### 2.2.8 External Evaluation of Lacuna

Before focusing on the assessment of the run-time overhead of JavaScript dead code, it is important to be reasonably confident that Lacuna is the right instrument for the detection and removal of JavaScript dead code. We carried out a small-scale experiment to evaluate Lacuna against state-of-the-art tools that are currently able to detect (and remove, in some cases) JavaScript dead code. In this section, we report the results of such an experiment.

We first identify an initial set of analysis tools that are currently able to detect JavaScript dead code. This step is carried out by: *(i)* performing a lightweight search on Google Scholar, and *(ii)* by analysing the scientific publications cited and citing the studies we already identified as related to our work (see Section 6). This activity leads to the following 6 promising tools: Qiong *et al.* [45], UFFRemover [33], JSLIM [46], Muzeel [47], Goel *et al.* [48], Google LightHouse[6]. Three researchers assessed the applicability of each potentially-usable tool (e.g., a functioning implementation of the tool must be publicly available). This analysis led to the identification of two tools that are usable in our study: UFFRemover and Muzeel. For the sake of space, the details of such analysis are included in the replication package of the study. The main distinguishing factors of the selected tools with respect to Lacuna are: *(i)* both UFFRemover and Muzeel detect dead code via dynamic analysis, whereas Lacuna can combine static and dynamic analyses; *(ii)* UFFRemover performs a preliminary static analysis to identify required JavaScript modules and to instrument them for logging the JavaScript functions executed during the dynamic analysis; *(iii)* the dynamic analysis of UFFRemover can execute various parts of the web app under analysis by executing test cases (if available) or via (user-defined) interaction scripts; *(iv)* Muzeel complements the initial loading of the web app with the emulation of all possible interactions within the web app (interaction points are identified during a preliminary pass via dynamic analysis); *(v)* Lacuna is a meta-tool, i.e., it allows the integration of additional 3rd-party analyzers in its pipeline; and *(vi)* Lacuna is the only tool supporting different optimization levels, where one of them – level 1 – is the one provided by UFFRemover [33].

We execute the UFFRemover and Muzeel tools on all the 39 TodoMVC web apps we used for the internal evaluation of Lacuna; for the sake of completeness, we execute two different configurations of UFFRemover, where the first one focusses exclusively on the initial load of the analysed web app (we call it *UFFRemover (L)*) and the second one is considering (scripted) interaction scenarios covering all functionalities of the analysed web app (we call it *UFFRemover (I)*). Finally, we consider the outputs of the three tools (i.e., Muzeel, UFFRemover (L), and UFFRemover (I)) over all 39 TodoMVC apps, we compute their precision, recall, and F-score, and finally we compare them against the same metrics we collected for the Dynamic-TAJS instance of Lacuna (see Section 2.2.7).

TABLE 2: Descriptive statistics for the external evaluation of Lacuna against Muzeel, UFFRem. (L), and UFFRem. (I).

| Tool | Min. | Max. | Median | Mean | SD | CV |
|---|---|---|---|---|---|---|
| **Precision** | | | | | | |
| Lacuna | 0.207 | 0.992 | 0.870 | 0.825 | 0.181 | 21.988 |
| Muzeel | 0 | | | 0.632 | 0.346 | 54.763 |
| UFFRem. (L) | 0.200 | 1 | 0.965 | 0.877 | 0.178 | 20.313 |
| UFFRem. (I) | 0.421 | 1 | 1 | 0.949 | 0.143 | 15.113 |
| **Recall** | | | | | | |
| Lacuna | 0.688 | 1 | 1 | 0.972 | 0.065 | 6.721 |
| Muzeel | 0 | 1 | 0.749 | 0.685 | 0.216 | 31.552 |
| UFFRem. (L) | 0.053 | 1 | 1 | 0.833 | 0.287 | 34.494 |
| UFFRem. (I) | 0.053 | 1 | 0.975 | 0.791 | 0.302 | 38.146 |
| **F-score** | | | | | | |
| Lacuna | 0.344 | 0.996 | 0.918 | 0.879 | 0.138 | 15.655 |
| Muzeel | 0 | | 0.714 | 0.594 | 0.302 | 50.831 |
| UFFRem. (L) | 0.101 | 1 | 0.891 | 0.801 | 0.247 | 30.904 |
| UFFRem. (I) | 0.101 | 1 | 0.955 | 0.810 | 0.255 | 31.430 |

https://developer.chrome.com/docs/lighthouse/performance/unused-javascript

Table 2 shows the descriptive statistics for the external evaluation of Lacuna against Muzeel, UFFRemover (L), and UFFRemover (I). UFFRemover (I) is the tool with the highest **precision** (mean=0.949). This result is expected since the interaction scripts we developed for interacting with the subjects are covering all primary functionalities of the analysed TodoMVC web apps; this result is also highlighted by the fact that the median precision of UFFRemover (I) is 1, i.e., the tool correctly identifies *all* dead functions as dead for at least 50% of the subjects. UFFRemover (L) (mean=0.877) and Lacuna (mean=0.825) perform similarly in terms of average precision; we conjecture that this result is primarily due to the fact that UFFRemover's detection algorithm executed on only the page loading phase of the subject is the same as the Dynamic analyzer of Lacuna (we trace the small difference in terms of precision to the fact that the two tools use a different library for parsing the JavaScript code, which might have lead to some functions not been detected by the parser).

Lacuna is the tool with the highest **recall** (mean=0.972), followed by UFFRemover (L) (mean=0.833), UFFRemover (I) (mean=0.791), and Muzeel (mean=0.685). As described in Section 2.2.7, having a high recall is fundamental for our experiment on its overhead at run-time (see Section 3) since a high recall makes us reasonably confident that (on average) Lacuna is able to detect 97.2% of all dead functions in a given web app. We conjecture that Lacuna is performing better than all the other tools since our Dynamic-TAJS instance of Lacuna includes also a static analysis component in it, allowing our tool to reach parts of the Javascript call graph that is not reached via either *(i)* the pure dynamic analysis performed by UFFRemover (L) and UFFRemover (I) or *(ii)* the dynamic analysis combined with the traversal of the event listeners statically-identified by Muzeel.

When looking at the F-score combined metric, Lacuna is again the tool performing better (mean=0.879), followed by UFFRemover (I) (mean=0.810), UFFRemover (L) (mean=0.801), and finally Muzeel (mean=0.594). The fact that Lacuna is the most accurate tool overall (i.e., it has the highest F-score) makes us reasonably confident in using it for detecting and removing JavaScript dead code in the subjects used when assessing the run-time overhead of JavaScript dead code (see next section).

## 3 EXPERIMENT ON THE RUN-TIME OVERHEAD OF JAVASCRIPT DEAD CODE

In this section, we describe the main aspects of the design of the experiment on the run-time overhead of JavaScript dead code. This experiment has been designed and conducted by following well-known guidelines for experimentation and data analysis in empirical software engineering [49], [50], [51], [52], [53]. We refer the reader to the replication package of the experiment [14] for further details on the experiment execution, used tools, and collected data. The replication package contains all the information for independent verification and replication of the study, namely: *(i)* the Python scripts for executing the experiment, *(ii)* the raw data measures collected during the execution of the experiment, and *(iii)* the R scripts for analysing the collected data, and *(iv)* a detailed guide for replicating the experiment.

### 3.1 Goal and Research Questions

In this context, we use Lacuna to eliminate dead code from the subjects of the experiment according to the four optimization levels of Lacuna (see Section 2.2.4). By following the GQM (Goal-Question-Metric) template [54], the goal of this experiment is formulated as:

> *Analyze the presence of JavaScript dead code **for the purpose of** empirically assessing its run-time overhead **with respect to** energy consumption, performance, network usage, and resources usage **from the point of view of** researchers, developers, and users **in the context of** mobile web apps.*

The goal presented above is achieved by answering the four research questions listed below. The main motivation for having the four research questions is to investigate the overall overhead that JavaScript dead code can have on mobile web apps at run-time. We define a research question for each of the main perspectives under which having a run-time overhead might be relevant either for the user experience or for the (technical, ecological) sustainability of mobile web apps.

> **RQ1.** What is the overhead of JavaScript dead code on the *energy consumption* of mobile web apps?

It is known that mobile web apps consume different amounts of energy while being loaded [55], [56] and that improving their energy efficiency might lead to consistent savings in terms of electricity [57]. So, answering RQ1 will help both web developers and researchers understand to what extent removing JavaScript dead code might be a useful instrument for improving mobile web apps from the perspective of energy consumption.

> **RQ2.** What is the overhead of JavaScript dead code on the *performance* of mobile web apps?

For what concerns RQ2, the performance of mobile web apps is a crucial factor for their success. Users expect mobile web apps to load within a reasonable time [58]; having mobile web apps with poor performance can potentially impact profits and/or lead to users' abandonment, especially on mobile devices where hardware and connectivity are constrained [59]. By answering RQ2 we aim to objectively assess to what extent the removal of JavaScript dead code might support *(i)* web developers in improving the performance of their mobile web apps and *(ii)* researchers in better understanding the relationship between the presence of (dead) JavaScript code and the performance of mobile web apps.

> **RQ3.** What is the overhead of JavaScript dead code on the *network usage* of mobile web apps?

It has been empirically confirmed that networking is the most relevant bottleneck for mobile web apps [5]. Also, the network conditions under which a mobile device operates can be limited depending on factors such as the network coverage at a specific location, the connectivity subscription of the user, the type of cellular network supported by the mobile device (e.g., 4G, 5G), etc. So, reducing the amount of network traffic required by a mobile web app to fully load is a relevant factor for improving its performance or even its loading itself. By answering RQ3 we aim at getting empirical evidence about the impact of JavaScript dead code and the network traffic required to load a mobile web app. Such results support both web developers and researchers in understanding if removing JavaScript dead code is a viable tool for reducing the requirements of mobile web apps in terms of network traffic.

**RQ4.** What is the overhead of JavaScript dead code on the *resources usage* of mobile web apps?

Mobile devices tend to have limited hardware resources, such as CPUs, GPUs, and memory. Also, the browser engine shares such resources with other apps running on the user's device and, when such resources are getting abused, the device might become slow and the operating system might even decide to forcibly shutdown some of the running apps to free resources for the other ones. By answering RQ4 we aim to empirically assess to what extent the presence of JavaScript dead code impacts the usage of hardware resources of the mobile device. Our results can support web developers and researchers in understanding if removing JavaScript dead code might help to reduce the number of resources needed to run a mobile web app, thus leading to an overall better user experience for their users.

### 3.2 Subjects Selection and Planning

For this experiment, we consider a total of 30 web apps that have been independently developed by third-party web developers. The 30 web apps are divided into two different families: 15 *in-the-lab* web apps and 15 *in-the-wild* web apps.

The 15 in-the-lab subjects were randomly sampled from the TodoMVC project. This project aims to help developers to choose the MV* framework more suitable for structuring and organizing their JavaScript Web apps. To that end, TodoMVC consists of different implementations of the "same" Todo web app, each of which uses a different MV* framework, so that developers can inspect the codebase and then compare the different MV* frameworks. The Todo app is a manager for to-do lists, which includes the following features: *(i)* adding a to-do item, *(ii)* removing a to-do item, *(iii)* modifying an existing to-do item, and *(iv)* marking a to-do item as completed. We refer to each sampled web app as the name of the used MV* framework. The sampled in-the-lab subjects are listed in Table 3.

Despite the in-the-lab web apps allowing us to study a *large* and *heterogeneous* set of MV* frameworks that real-world JavaScript Web apps can rely on, they share the same functionalities. This might negatively influence the external validity of the experiment, making our results less generalizable. In order to mitigate this potential bias, we decided to complement the 15 in-the-lab subjects with 15 additional in-the-wild subjects. The subjects are sampled from the Tranco list [12], which aggregates the rankings from the lists provided by Alexa, Umbrella, Majestic, and Quantcast. Starting from the first $150$ web apps in the Tranco list, we iteratively download and manually analyze each candidate web app against a set of selection criteria we defined a priori, reaching a final set of 15 web apps satisfying all the selection criteria. The selection criteria, their rationale, and the results of their application are reported below:

S1 – *The web app should not redirect to another domain*. The rationale for this criterion is that there are mobile web apps that redirect the user to a different domain, such as Apple (aaplimg.com → apple.com) and Twitter (t.co → twitter.com); these pages could redirect to duplicate domains within the list or domains that are not part of the Tranco list at all. The application of S1 led to the identification of 24 web apps redirecting to another domain, which were discarded from the initial 150 web apps.

S2 – *The web app must be accessible without user authentication*. The rationale for this criterion is that there are mobile web apps in which the actual page content is available for authenticated users only, such as Twitter and Instagram. After applying this criterion we identified 8 web apps requiring user authentication, leading to a set of 118 potentially-usable web apps.

S3 – *Lacuna must be able to successfully remove JavaScript dead code from the web app without errors*. The rationale for this criterion is that we need to be sure that for each subject of the experiment, we can successfully run Lacuna to obtain its dead-code-free version for all Lacuna optimization levels. When applying Lacuna to the 118 selected web apps we encountered two main situations where it was not successful: *(i)* 96 web apps included external JavaScript scripts we did not manage to properly download locally on our server (i.e., some scripts were imported dynamically and the browser blocked their request due to Cross-Origin Resource Sharing errors, the HTML code of the web app was referencing scripts which were not available anymore at the referenced URLs, etc.) and *(ii)* for 7 web apps TAJS failed since at the time of executing the experiment it did not support the following ES6 features: the let keyword, arrow functions, and template literals.

The 15 selected in-the-wild subjects resulting from this procedure are listed in Table 3. These subjects are heterogeneous from different perspectives (e.g., application domain, functionalities, size, amount of JavaScript code), making them good candidates for complementing the in-the-lab subjects and achieving more generalizable results in our experiment.

Once we obtain the final set of 30 individual subjects, we apply Lacuna four times to each of them, each time with a different optimization level (OL-0 as the baseline, then OL-1, OL-2, OL-3). This leads to have four versions for each web apps. In Table 3, we report the number of functions detected as dead by Lacuna when it is executed on each web app—it is worth recalling that such a number is the same across the different Lacuna optimization levels.

Regardless of the web app and optimization level, we configured Lacuna so that it combined the results of two third-party analysis techniques: Dynamic and TAJS. This design choice was taken empirically. That is, we performed a preparatory experiment thanks to which we concluded that the best configuration of Lacuna was the one based on the joint use of Dynamic and TAJS (see Section 2.2.7).

### 3.3 Variables and Statistical Hypotheses

This experiment has the same **independent variable** for all research questions, i.e., the Lacuna *optimization level* applied to each of the subjects. According to the currently available dead code elimination procedure of Lacuna described in Section 2.2.4, this variable has four levels: OL-0, OL-1, OL-2, and OL-4.

TABLE 3: Number of dead functions detected by Lacuna for each subject.

| Subject | # Dead Functions |
|---|---|
| **In-the-lab subjects** | |
| angularjs require | 32 |
| backbone | 542 |
| canjs | 492 |
| dijon | 410 |
| dojo | 411 |
| enyo backbone | 6 |
| gwt | 17 |
| jquery | 420 |
| jsblocks | 459 |
| knockoutjs require | 35 |
| mithril | 55 |
| polymer | 6 |
| reagent | 3,357 |
| vanillajs | 59 |
| vue | 266 |
| **In-the-wild subjects** | |
| apache.org | 437 |
| aws.amazon.com | 409 |
| m.youtube.com | 1,812 |
| nl.godaddy.com | 19 |
| stackexchange.com | 457 |
| stackoverflow.com | 491 |
| www.amazon.com | 144 |
| www.bbc.com | 345 |
| www.booking.com | 1,390 |
| www.buzzfeed.com | 353 |
| www.mozilla.org | 436 |
| www.office.com | 616 |
| www.paypal.com | 639 |
| www.theguardian.com | 16 |
| www.wikipedia.org | 46 |

All **dependent variables** are measured in the time frame between the first GET request issued by the browser to the server hosting the currently-measured web app and the web app's page load time. The dependent variables of this experiment are described below:

- *Energy* (RQ1): the energy consumed by the mobile device to load the web app in mJ (milli-joule). Energy values are computed by following a sampling-based approach widely used in software engineering studies [60], [61], [62], [63], that is: *(i)* sampling the instantaneous power consumed by the browser app running on the Android device (in microWatts) *(ii)* applying the $E = P \times t$ formula, where $P$ is the measured power and $t$ is the page load time of the web app (see next dependent variable), and *(iii)* solving the integral of $P$ over $t$ (in our case by applying the trapezoidal method [64]).

- *Page load time* (RQ2): the timestamp in milliseconds (ms) in which the web app is fully loaded in the browser [65]. More specifically, page load time is defined as the time from the start of a user-initiated page request (the initial GET request issued by the browser in our case) to the time the entire page content is loaded, including all dependent resources like CSS stylesheets, JavaScript code, or images; this time is collected by recording the timestamp in which the load event is fired by the browser engine.

- *First contentful paint* (RQ2): the timestamp in milliseconds when the browser first renders any text,

image, non-white canvas, or SVG of the web app [65]. Intuitively, it is the first time when the user can start consuming the content of the web app. According to the Paint Timing W3C specification [66], the First contentful paint metric and the First paint one (see below) complement Page load time since they provide a user-oriented assessment of the performance of the web app.

- *First paint* (RQ2): the timestamp in milliseconds when the browser renders the first pixels to the screen of the mobile device, rendering anything that is visually different from what was on the screen prior to navigation [65]. Intuitively, it is the time when the user is aware that "something is happening" in the browser after they decided to navigate to the URL of the mobile web app.

- *HTTP requests* (RQ3): the number of HTTP(S) requests issued by the browser engine while loading the currently-measured web app. We include this variable since our RQ3 is concerning the overhead of JavaScript dead code in terms of network usage, mainly due to the additional network traffic caused by either the additional JavaScript files retrieved by the web app (even if they are not executed since they contain dead code).

- *Transferred bytes* (RQ3): the sum of the size, in kilobytes (Kb), of the payloads of all HTTP(S) requests issued by the browser engine while loading the currently-measured web app. Similarly to the previous variable, we are measuring the number of transferred bytes in order to quantify how much additional (and unused) JavaScript code is transferred from the servers to the web app when dealing with JavaScript dead code.

- *CPU usage* (RQ4): the average of the percentage of CPU consumed while loading the currently-measured web app. We include this variable since RQ4 deals with the overhead imposed by JavaScript dead code in terms of computational resources, which are typically the processor, GPU, and memory (see the description of the next two dependent variables).

- *GPU usage* (RQ4): the average of the percentage of GPU consumed while loading the currently-measured web app. Similarly to the previous variable, we include this variable in order to measure what is the added overhead of JavaScript dead code in terms of GPU usage.

- *Memory usage* (RQ4): the average amount of memory consumed by the Android device while loading the currently-measured web app in megabytes (Mb). Similarly to CPU usage, we include this variable in order to measure what is the memory overhead imposed by JavaScript dead code.

For each dependent variable listed above and each family of subjects (i.e., in-the-lab vs. in-the-wild ones), we formulate the following parameterized null hypothesis:

$H0_{var}$ : There is no statistically significant difference in the values of the dependent variable *var* (e.g.,

energy, page load time, etc.) between the optimization levels of Lacuna.

The alternative hypothesis for $H0_{var}$ (i.e., $H1_{var}$) admits that there is a statistically significant difference. For example, if $H0_{energy}$ is rejected, we can accept the alternative hypothesis $H1_{energy}$ stating that *there is a statistically significant difference in the values of energy between the optimization levels of Lacuna*.

### 3.4 Experiment execution

In Figure 4, we present the measurement infrastructure for running the experiment. The experiment involves two main hardware nodes: a laptop acting as a base station and an Android smartphone for running and measuring the subjects. The laptop has an Intel Core i7-4710HQ processor, 12Gb of memory, and runs Ubuntu 20.04 as the operating system. The Android device is an LG G2 smartphone with a Qualcomm MSM8974 Snapdragon 800 processor, 2Gb of memory, a 5.2" LCD display, and running Android 6.0.1 operating system. The main rationale for using two separate hardware nodes is to keep the Android device as lightweight as possible, so as to not influence the measurements [67], [68]. As shown in the right-hand side of the figure, the Android device is running only two apps: *(i)* the Google Chrome browser, which is used for loading the web apps and *(ii)* Trepn, a software-based profiler for Android devices. Trepn is widely used in empirical studies on energy-efficient software [69], [70], [71] and it has been reported as sufficiently accurate with respect to hardware power measurement (e.g., the Monsoon Power Monitor), with an error margin of 99% [72]. Trepn supports also the collection of the CPU, GPU, and memory usage.

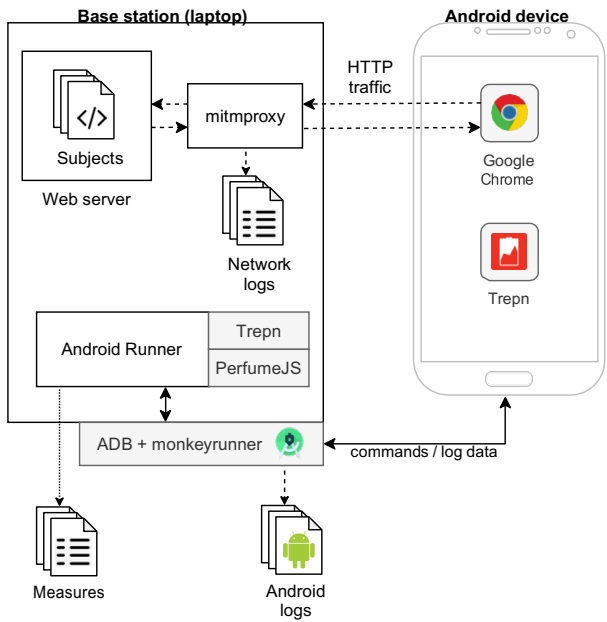

**Fig. 4: Measurement infrastructure**

The laptop and the Android device are connected to the same WiFi network. To reduce as much as possible the influence of the network conditions on the experiment, the WiFi network does not have any other connected devices.

All four versions of each of the 30 subjects of the experiment are hosted on the laptop and served via a dedicated Web server. To collect the values of the HTTP requests and Transferred bytes dependent variables, all HTTP(S) traffic between the smartphone and the laptop passes through an instance of *mitmproxy* [73], which records all HTTP(S) requests and locally stores them the form of network logs.

The experiment is orchestrated via Android Runner, a framework for defining and executing measurement-based experiments targeting Android (web) apps [67]. Android Runner allows us to define the experiment in a descriptive manner via a JSON file and then it automatically takes care of the complete execution of the experiment. Specifically, for each experiment run, Android Runner uses the Android Debug Bridge tool (ADB [74]) to interact with the smartphone, e.g., to collect Android system logs, to activate/deactivate the profiling features of Trepn, to instruct the Google Chrome app on the smartphone to load the currently-measured subject, to enable/disable the USB charging of the smartphone, etc.. For this experiment, we use two plugins of the Android Runner tool: *(i)* Trepn, for collecting data via the Trepn profiler and *(ii)* PerfumeJS, to collect web performance metrics via the Perfume.js library [75], such as the page load time, first contentful paint, and first paint.

In order to mitigate possible threats to the internal validity of the experiment and to facilitate its replicability, we take the following precautions while executing it: *(i)* the measurement of each experiment trial (i.e., a subject-OL pair) is repeated 20 times, leading to a total of 2,400 individual runs (i.e., 4 treatments x 30 subjects x 20 repetitions), *(ii)* the order of execution of the 2,400 experiment runs is randomized, *(iii)* between each run the smartphone and the laptop remain idle for 2 minutes so to take into account tail energy usage [76], *(iv)* the Chrome app is cleared before each run so to reset its cache, persisted data, and configuration, and *(v)* the USB charging of the smartphone is disabled during the execution of each run.

### 3.5 Data Analysis

We first perform a *data exploration* step where we inspect and get an overview of the collected data via box plots and summary tables. In this step, we also check if the assumption that in-the-lab and in-the-wild subjects exhibit different values holds for the considered metrics. Since normality of the collected data is the underlying assumption of parametric statistical tests [49], as part of the data exploration step, we check if the distribution of the data collected for each dependent variable follows a normal distribution, both globally and between in-the-lab and in-the-wild subjects. We assess normality by means of three complementary methods: *(i)* by applying the Shapiro-Wilk statistical test with $\alpha = 0.05$, *(ii)* by producing and visually analysing the density plots of every dependent variable, and *(iii)* by producing and visually analysing QQ-plots.

We anticipate that all the collected data do not follow a normal distribution. Based on this fact, in our statistical analysis, we apply non-parametric statistical tests and effect size measures. Specifically, *for each dependent variable and each family of subjects* (i.e., in-the-lab and in-the-wild) we do the following:

1) We apply the Kruskal-Wallis test (with $\alpha = 0.05$), a non-parametric test for testing whether the collected measures come from populations with identical distributions; the application of this test gives an initial indication about whether the Lacuna optimization levels lead to statistically-different differences in terms of, e.g., energy consumption, memory usage, etc.

2) If the p-value of the Kruskal-Wallis test is not greater than $\alpha$, then we assess the magnitude of the detected differences by applying the Eta squared effect size measure based on the H-statistic [77]. Eta squared is a non-parametric effect size measure compatible with the Kruskal-Wallis statistical test [77]. The values of the obtained effect size measures are interpreted according to threshold values commonly used in the literature, namely: $\eta^2 < 0.06$ (small effect - S), $0.06 \leq \eta^2 < 0.14$ (moderate effect - M), and $\eta^2 \geq 0.14$ (large effect - L).

3) Having a statistically significant result for the Kruskal-Wallis test also allows us to investigate which *pairs* of optimization level exhibit statistically-significant differences. We do so by applying the Dunn Test as post-hoc analysis [78] to each pair of optimization level. Since we are applying multiple statistical tests, to reduce the chance of Type-I error we adjust the obtained p-values via the Benjamini-Hochberg correction [79].

# 4 RESULTS

In this section, we present the results of the experiment.

## 4.1 Data exploration

Table 4 gives an overview of the measures we collected for all dependent variables across all in-the-wild and in-the-lab subjects. We observe that the measures collected for in-the-lab subjects tend to have different central values (i.e., mean and median) with respect to those collected from in-the-wild subjects; this phenomenon is especially prominent for page load time, first contentful paint, first paint, HTTP requests, transferred bytes, CPU usage, and memory usage. Also, when the central values are different, in-the-wild subjects tend to consistently perform worse with respect to in-the-lab subjects; for example, the average page load time of in-the-wild subjects is 4.229s, whereas it is 1.137s for in-the-lab subjects. The differences in the obtained measures for in-the-lab and in-the-wild subjects further validate our design choice of considering the type of subject as a blocking factor for our experiment. Indeed, we expected such a kind of difference since the purpose and context in which those two families of subjects are developed are completely different; in-the-lab subjects have a relatively small size (both in terms of provided features and source code) and are developed on a voluntary basis, whereas in-the-wild subjects are fully-fledged web apps developed either by *(i)* companies like Google or Amazon or *(ii)* large-scale organizations like the Wikimedia Foundation.

The collected data for each dependent variable exhibits values within the expected ranges. For example, energy consumption is between 343.39mJ and 1,819.67mJ, which are acceptable values if we consider that the average page load time of the measured subjects is relatively short (i.e.,

2.6 seconds). Overall, there is a high relative variance in the data, especially for network-related variables (i.e., HTTP requests and transferred bytes) for in-the-wild subjects and memory usage for all subjects, but also for performance-related metrics (i.e., page load time, first contentful paint, and first paint) for in-the-wild subjects. Such variance is not a surprise if we consider that the time span in which measures are collected is relatively short (it raises the chances of having outlier values) and that we are including in-the-wild subjects in the experiment.

TABLE 4: Descriptive statistics of the collected measures for all dependent variables (SD=standard deviation, CV=coefficient of variation)

| Variable | Min. | Max. | Median | Mean | SD | CV |
|---|---|---|---|---|---|---|
| **All subjects** | | | | | | |
| Energy (mJ) | 343.49 | 1,819.67 | 1,379.37 | 1,387.88 | 104.07 | 7.50 |
| Page load time (ms) | 454.00 | 17,857.00 | 1,897.00 | 2,683.57 | 2,425.06 | 90.37 |
| First cont. paint (ms) | 629.90 | 7,623.20 | 1,484 | 1,584.67 | 764.29 | 48.23 |
| First paint (ms) | 561.60 | 7,623.20 | 1,241.35 | 1,453.87 | 734.56 | 50.52 |
| HTTP requests | 5 | 102 | 13 | 20.27 | 18.15 | 89.52 |
| Transferred bytes (Kb) | 35.26 | 2,535.35 | 291.58 | 415.70 | 412.02 | 99.11 |
| CPU usage (%) | 52.18 | 93.14 | 59.06 | 59.73 | 3.41 | 5.71 |
| GPU usage (%) | 10.14 | 35.34 | 24.1 | 23.86 | 2.77 | 11.59 |
| Memory usage (Mb) | 1,580.40 | 1,766.24 | 1,668.47 | 1,669.63 | 30.45 | 1.82 |
| **In-the-lab subjects** | | | | | | |
| Energy (mJ) | 1,203.56 | | | 1,386.13 | 102.04 | 7.36 |
| Page load time (ms) | 454.00 | 12,357 | 837.00 | 1,137.36 | 835.32 | 73.44 |
| First paint (ms) | 561.60 | 4,455 | 923.40 | 969.26 | 372.22 | 38.40 |
| HTTP requests | 8 | 42 | 14 | 15.87 | 7.76 | 48.91 |
| Transferred bytes (Kb) | 35.26 | 2,535.35 | 212.56 | 398.77 | 496.67 | 124.55 |
| CPU usage (%) | 52.18 | 71.43 | 58.41 | 58.58 | 1.69 | 2.88 |
| GPU usage (%) | 14.44 | 30.75 | 24.81 | 24.57 | 2.53 | 10.31 |
| Memory usage (Mb) | 1,580.40 | 1,766.24 | 1,657.84 | 1,658.74 | 26.82 | 1.62 |
| **In-the-wild subjects** | | | | | | |
| Energy (mJ) | 343.49 | 1,749.92 | 1,381.92 | 1,389.63 | 106.07 | 7.63 |
| Page load time (ms) | 1,045.00 | 17,857.00 | 3,660.50 | 4,229.78 | 2,506.69 | 59.26 |
| First cont. paint (ms) | 718.70 | 7,623.20 | 1,849.95 | 2,012.69 | 649.40 | 32.27 |
| First paint (ms) | 730.50 | 7,623.20 | 1,799.70 | 1,938.49 | 686.28 | 35.40 |
| HTTP requests | 5 | 102 | 13 | 24.67 | 23.66 | 95.91 |
| Transferred bytes (Kb) | 100.43 | 1,353.06 | 342.10 | 432.64 | 304 | 70.27 |
| CPU usage (%) | 53.88 | 93.14 | 60.02 | 60.89 | 4.21 | 6.91 |
| GPU usage (%) | 10.14 | 35.34 | 23.43 | 23.15 | 2.81 | 12.12 |
| Memory usage (Mb) | 1,591.47 | 1,704.23 | 1,676.19 | 1,680.53 | 29.90 | 1.78 |

The Shapiro-Wilk normality test reveals that all the data exhibit a non-normal distribution. This result is further confirmed visually via density plots and QQ-plots (all available in the replication package of this study). Since the normality of the data is one of the assumptions of the ANOVA statistical test, we resort to the Kruskal-Wallis test as a non-parametric statistical test in the remainder of our data analysis procedure.

In the next sections, we analyze the data related to each research question of the experiment.

## 4.2 Overhead on Energy Consumption (RQ1)

As shown on the left-hand side of Figure 5, eliminating JavaScript dead code from the in-the-lab subjects results in slightly more energy-efficient web apps. Indeed, the median energy consumption of the original web apps (OL-0 in Lacuna) is 1,406.35mJ, against 1,367.59mJ, 1,374.34mJ, and 1,370.83mJ for the other Lacuna optimization levels. However, this result is not statistically significant (p-value: 0.268, see the first row of Table 5).

TABLE 5: Results of the statistical analysis for all RQs – the (*) symbol denotes cases with statistically-significant differences among optimization levels

| Variable | Subject type | P-value | Effect size |
|---|---|---|---|
| **Overhead on energy consumption (RQ1)** | | | |
| Energy (mJ) | Lab | 0.268 | - |
| | Wild | 0.768 | - |
| **Overhead on performance (RQ2)** | | | |
| Page load time (ms) | Lab | $4.7 \times 10^{-9}$ (*) | 0.032 (S) |
| | Wild | 0.039 (*) | 0.004 (S) |
| First cont. paint (ms) | Lab | 0.654 | - |
| | Wild | 0.428 | - |
| First paint (ms) | Lab | 0.461 | - |
| | Wild | 0.907 | - |
| **Overhead on network usage (RQ3)** | | | |
| HTTP requests | Lab | $1.48 \times 10^{-40}$ (*) | 0.155 (S) |
| | Wild | 0.937 | - |
| Transferred bytes (Kb) | Lab | $2 \times 10^{-14}$ (*) | 0.054 (S) |
| | Wild | $2.19 \times 10^{-20}$ (*) | 0.077 (M) |
| **Overhead on resources usage (RQ4)** | | | |
| CPU usage (%) | Lab | 0.0552 | - |
| | Wild | $135 \times 10^{-20}$ (*) | 0.027 (S) |
| GPU usage (%) | Lab | 0.002 (*) | 0.011 (S) |
| | Wild | 0.206 | - |
| Memory usage (Mb) | Lab | 0.337 | - |
| | Wild | 0.0159 (*) | 0.006 (S) |

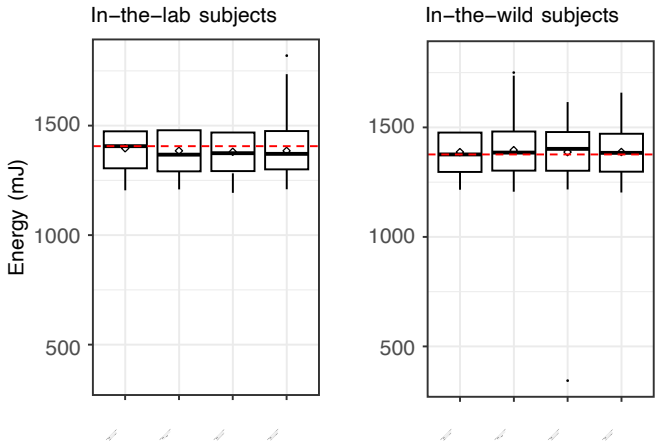

Fig. 5: Energy consumed by the mobile device (the red dashed line represents the median value for the original web apps (OL-0))

The results are similar for in-the-wild subjects, with the exception that the median energy consumption remains approximately the same for the OL-1 and OL-3 Lacuna optimization levels. The energy consumption of the OL-2 optimization level is even slightly higher than that of the others, but still without statistical significance; we speculate that this result is due to the intrinsic variability of the experiment execution.

Overall, the obtained p-values do not allow us to reject $H0_{energy}$ for both in-the-lab and in-the-wild subjects, so we cannot claim that different Lacuna optimization levels have an impact on the energy consumption of in-the-lab or in-the-wild web apps.

## 4.3 Overhead on Performance (RQ2)

Eliminating JavaScript dead code from the web apps leads to an improvement in the page load time (see Figure 6). This result is statistically significant for both in-the-lab and in-the-wild subjects, with a p-value of $4.7 \times 10^{-9}$ for in-the-lab subjects and a p-value of 0.039 for in-the-wild subjects (see Table 5). Given the obtained p-values, we can reject $H0_{page\ load\ time}$ for both families of web apps, allowing us to claim that different Lacuna optimization levels have an impact on the page load time of web apps (both in-the-lab and in-the-wild). However, our effect size estimation reveals that the effect of JavaScript dead code elimination on page load time is *small* (S) for both of them (see Table 5).

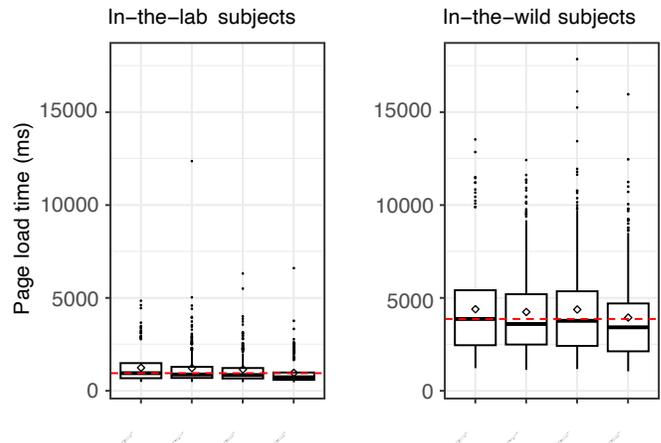

Fig. 6: Page load time of the web apps

Having a statistically-significant result for the Kruskal-Wallis test on page load times means that we can investigate which Lacuna optimization level leads to statistically-significant differences with respect to the original web app and the other optimization levels. As shown in the first two rows in Table 6, completely eliminating the JavaScript functions identified as dead code (i.e., OL-3) leads to a statistically-significant difference for both in-the-lab and in-the-wild subjects. This difference is also visually highlighted in the box plots in Figure 6, where OL-3 tends to have lower values with respect to the others. Other statistically-significant pairs of optimization levels are OL-1 → OL-3 and OL-2 → OL-3 for in-the-lab subjects; this result can be seen as an indication that, in the context of in-the-lab subjects, having an approach with different strategies for eliminating JavaScript dead code paid off in terms of page load time and that having a more comprehensive (but risky) strategy for dead code elimination leads to significantly better results in terms of page load time.

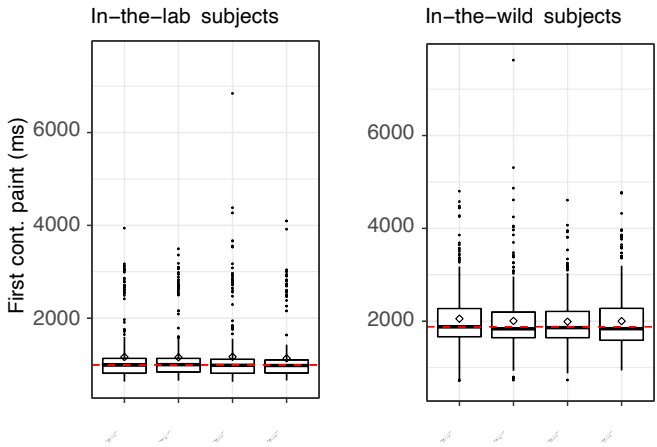

Fig. 7: First contentful paint of the web apps

Differently from page load time, the measures of first contentful paint (Figure 7) and first paint (Figure 8) do not show any noticeable improvement across the various Lacuna optimization levels. As shown in Table 5, we obtained p-values higher than our significance threshold for both first contentful paint and first paint, thus we cannot reject the corresponding null hypotheses for both families of subjects. That is, we cannot claim that different Lacuna optimization levels impact differently the time of the first contentful paint and first paint of web apps (both in-the-lab and in-the-wild subjects).

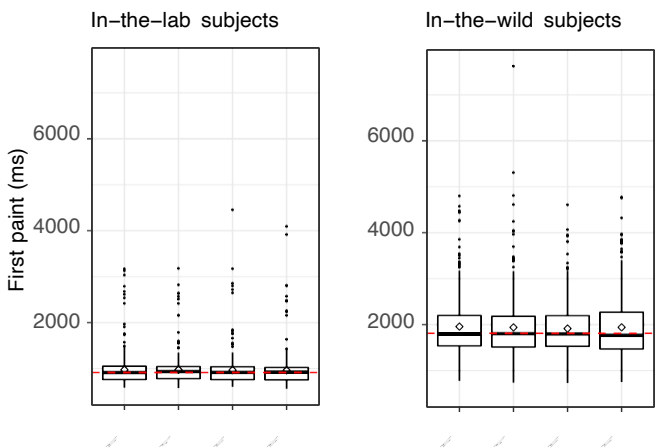

Fig. 8: First paint of the web apps

## 4.4 Overhead on Network Usage (RQ3)

As shown in Figure 9 and Table 5, the elimination of JavaScript dead code leads to noticeable (and statistically significant, p-value: $1.48 \times 10^{-40}$) differences in terms of the number of HTTP requests only for in-the-lab subjects. Moreover, by looking at Table 6, it can be noticed that such differences are statistically significant for all pairs involving the OL-3 optimization level of Lacuna. In any case, the observed effect size is 0.155, i.e., *small*. The situation is different for in-the-wild subjects, where we do not observe any relevant difference among the various Lacuna optimization levels. Summing up, only for in-the-lab subjects (not for in-the-wild ones) we can reject $H0_{HT\ T\ P\ requests}$ stating that the number of HTTP requests is not the same across different Lacuna optimization levels.

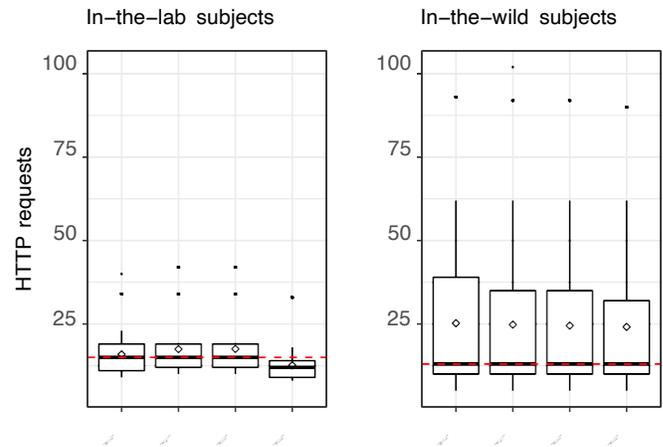

Fig. 9: Number of HTTP requests

The amount of transferred bytes is considerably lesser when the various Lacuna optimization levels are applied to both in-the-lab and in-the-wild subjects (see Figure 10). The differences in the transferred bytes are statistically significant for both in-the-lab subjects (p-value: $2 \times 10^{-14}$, effect size: 0.054 – *small*) and in-the-wild subjects (p-value: $2.19 \times 10^{-20}$, effect size: 0.077 – *moderate*). So, we can reject $H0_{transferred\ bytes}$ for both families of subjects, allowing us to claim that different Lacuna optimization levels have an impact on the transferred bytes from the server to the client web apps (both in-the-lab and in-the-wild). As shown in Table 6, the observed differences are statistically significant for almost all pairs of Lacuna optimization levels.

## 4.5 Overhead on Resources Usage (RQ4)

The CPU usage remains stable for in-the-lab subjects (left-hand side of Figure 11), with an average close to 58% and a p-value of 0.552. Differently, the CPU usage for in-the-wild subjects is reduced when applying Lacuna (p-value: $135 \times 10^{-9}$), with statistically-significant results for all pairs of optimization levels, but not for the OL-0_OL-1 one. The obtained p-values tell us that we cannot reject $H0_{cpu\ usage}$ for in-the-lab subjects, but we can reject such a null hypothesis for in-the-wild subjects—i.e., there is a difference in the percentage of CPU usage across different Lacuna optimization levels when considering in-the-wild subjects.

The scenario is more stable when considering GPU usage (see Figure 12, with an average close to 25% and 24% for in-the-lab and in-the-wild subjects, respectively. The row of Table 5), further confirmed for the OL-0→OL-3 and the OL-1→OL-3 pairs of optimization levels (see the last row of Table 6). Given the obtained p-values, we cannot reject $H0_{memory\ usage}$ for in-the-wild web apps (i.e., there is a difference in the memory usage of the web apps across different Lacuna optimization levels), but we can reject the same null hypothesis for in-the-lab web apps.

TABLE 6: Pairwise comparison across Lacuna optimization levels for variables with statistically-significant differences - the (*) symbol denotes cases with statistically-significant differences (with Benjamini-Hochberg correction)

| Variable | Subject type | RQ | OL-0 → OL-1 | OL-0 → OL-2 | OL-0 → OL-3 | OL-1 → OL-2 | OL-1 → OL-3 | OL-2 → OL-3 |
|---|---|---|---|---|---|---|---|---|
| Page load time (ms) | Lab | RQ2 | 0.367 | 0.119 | $1.21 \times 10^{-8}$ (*) | 0.464 | $1.97^{-6}$ (*) | $4.45 \times 10^{-4}$ (*) |
| Page load time (ms) | Wild | RQ2 | 0.691 | 0.737 | 0.049 (*) | 0.737 | 0.114 | 0.063 |
| HTTP requests | Lab | RQ3 | 0.205 | 0.163 | $8.97 \times 10^{-24}$ (*) | 0.813 | $4.48 \times 10^{-30}$ (*) | $5.63 \times 10^{-31}$ (*) |
| Transferred bytes (Kb) | Lab | RQ3 | 0.028 (*) | $1.67 \times 10^{-4}$ (*) | $1.45 \times 10^{-14}$ (*) | 0.109 | $4.66 \times 10^{-18}$ (*) | $1.01 \times 10^{-4}$ (*) |
| Transferred bytes (Kb) | Wild | RQ3 | $1.80 \times 10^{-4}$ (*) | $7.44 \times 10^{-14}$ (*) | $1.24 \times 10^{-17}$ (*) | $1.80 \times 10^{-4}$ (*) | $1.71 \times 10^{-6}$ (*) | 0.258 |
| CPU usage (%) | Wild | RQ4 | 0.462 | 0.031 (*) | $1.26 \times 10^{-5}$ (*) | $5.99 \times 10^{-3}$ (*) | $5.72 \times 10^{-7}$ (*) | 0.027 (*) |
| GPU usage (%) | Lab | RQ4 | 0.201 | 0.061 | 0.197 | 0.433 | 0.012 (*) | $1.44 \times 10^{-3}$ (*) |
| Memory usage (Mb) | Wild | RQ4 | 0.871 | 0.532 | 0.021 (*) | 0.534 | 0.021 (*) | 0.106 |

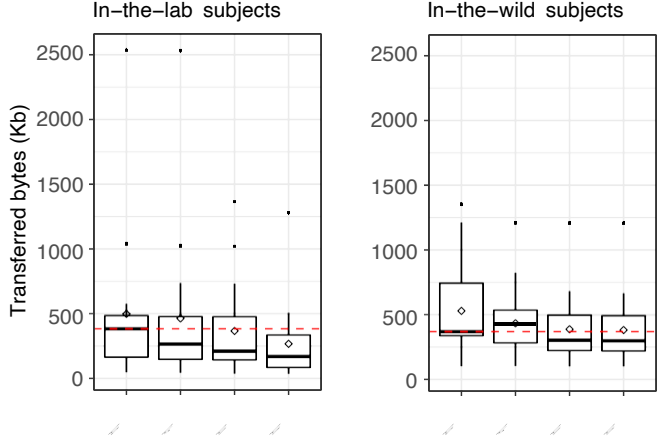

Fig. 10: Bytes transferred over the network

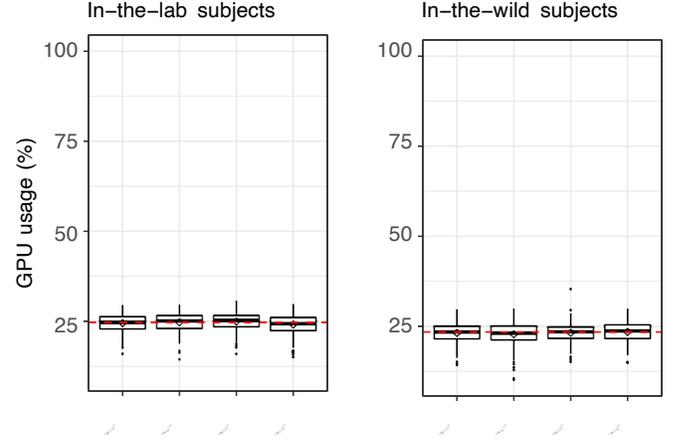

Fig. 12: GPU usage of the mobile device

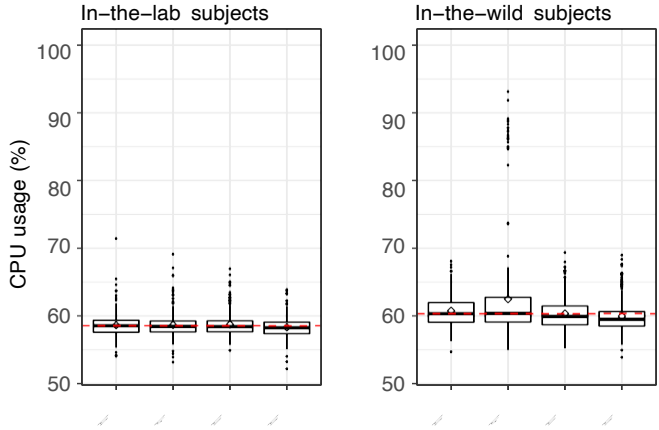

Fig. 11: CPU usage of the mobile device

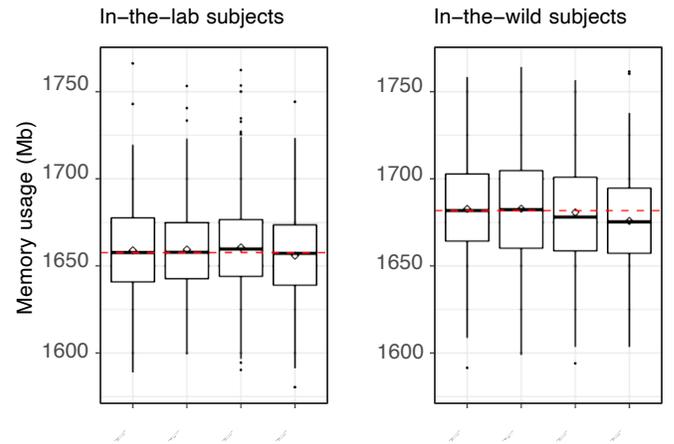

Fig. 13: Memory usage of the mobile device

application of the Kruskal-Wallis test reveals a statistically-significant difference only for in-the-lab subjects (p-value: 0.002 in Table 5) and the OL-1→OL-3 and the OL-2→OL-3 pairs of optimization levels. So, we can reject $H0_{gpu\ usage}$ for in-the-lab web apps (i.e., different Lacuna optimization levels impact differently the percentage of GPU usage), but we cannot reject the same hypothesis for in-the-wild ones.

Memory usage remains relatively stable for in-the-lab subjects (see Figure 13), whereas it exhibits an observable improvement when considering in-the-wild subjects, with a statistically-relevant difference (p-value: 0.0159 in the last

Despite the observed statistical differences, the effect size for CPU usage, GPU usage, and memory usage remains *small* (see Table 5).

## 4.6 Impact of Dead Function Removal

In the following, we present the results of a further analysis to study the potential correlations between the number of dead functions detected by Lacuna and the improvements due to the elimination of these functions (i.e., by applying OL-1, OL-2, OL-3, respectively) in terms of *energy consumption*, *performance*, *network usage*, and *resources usage*. To do so, we run a correlation test between the number of dead functions detected by Lacuna and the saving achieved after applying each Lacuna optimization level (except for OL-0) with respect to each measure listed in Section 3.3. In particular, given a measure and an optimization level (except for OL-0), the saving on each subject is computed by averaging the 20 measurements for OL-0 (i.e., no optimization) and then subtracting the average of the 20 measurements for the considered optimization level. We use the Kendall correlation coefficient because to apply such a test it is not required that the data are normally distributed. Moreover, by averaging the data, we meet the data independence assumption. If the p-value of the correlation test is not greater than $\alpha = 0.05$, then there is a statistically-significant correlation. In this case, we then report the Kendall correlation coefficient (Tau), which provides an indication of how strong a statistically-significant correlation is. The values of this correlation coefficient are interpreted according to threshold values commonly used in the literature, namely: $\tau < 0.1$ (no correlation), $0.1 \leq \tau < 0.4$ (weak correlation - W), $0.4 \leq \tau < 0.7$ (moderate correlation - M), and $\tau \geq 0.7$ (strong correlation - S).

In Table 7, we show the results of this further analysis. For most of the measures, we cannot show there are statistically-significant correlations with respect to the number of dead functions Lacuna detected. Among the most important outcomes, there is that in terms of network usage. In particular, we find positive, moderate, and statistically-significant correlations between the number of dead functions and the number of transferred bytes for all optimization levels (p-values ranging from $2.94 \times 10^{-6}$ to $4.11 \times 10^{-4}$, correlation coefficients ranging from 0.456 to 0.603). That is, regardless of the Lacuna optimization level, when the number of dead functions increases, the saving on transferred bytes tends to increase as well.

The other statistically-significant correlations regard the saving in terms of performance (i.e., page load time) and resources usage (i.e., CPU usage). In particular, we have a positive and statistically-significant correlation between the number of dead functions and saving on page load time when applying OL-3 (p-value: 0.009, correlation coefficient: $0.336 - weak$). That is, as the number of dead functions removed with OL-3 increases, the saving on page load time tends to increase. Finally, we find a positive, *weak*, and statistically-significant correlation between the number of dead functions and saving on CPU usage when applying OL-2 (p-value: 0.022, correlation coefficient: 0.295). That is, the more the number of dead functions removed after applying OL-2, the higher the saving on CPU usage.

## 5 DISCUSSION

In this section, we discuss the obtained results in terms of implications from the perspectives of users, researchers, and web developers. We conclude the section by presenting possible threats that could affect the validity of the obtained results, including countermeasures we applied for mitigating them.

## 5.1 Implications

We observed that **page load time** is significantly lower when JavaScript dead code is completely removed (i.e., Lacuna optimization level 3) whatever the family of web apps is. Taking into account that users tend to abandon a web app when it takes too much time to load pages [59], we can postulate that our outcome has practical implications from the perspective of *users*. For example, users could appreciate the page load time of mobile web apps without dead code, thus positively affecting the user experience[7] of these web apps. This outcome has also implications from the perspective of *researchers* because they could be interested in studying to what extent the user experience is affected by the presence, or not, of dead code in a given mobile web app. From the *web developer* perspective, it could be relevant to integrate a tool, like Lacuna, in the user experience design process since the presence of dead code could significantly affect page load time, possibly affecting user experience.

Regardless of the family of web apps, the amount of **transferred bytes** from the server to the client is significantly less when the various Lacuna optimization levels are applied, especially when using optimization level 3. That is, we provide evidence that removing JavaScript dead code is effective in reducing the network transfer of web apps. We can postulate that this outcome is relevant for *users* and *web developers* since transferring fewer bytes is definitely valuable when the network is either a scarce resource (e.g., due to low available bandwidth) or underpaid/limited subscription plans (i.e., for 4G/5G connectivity). *Web developers* could find this outcome interesting also because a small earn in the transferring of bytes from the Cloud to a single client reflects in a large earn (in terms of transferred bytes) when millions of clients are connected to the Cloud. *Researchers* could be interested in studying to what extent the mentioned earn affects the *end-to-end* energy consumption, i.e., the total energy needed to transfer bytes from the Cloud to the clients, including the consumption of network devices (e.g., switches and routers). This proposed line of research aims at deepening the results of RQ1, where we could not find evidence that the removal of JavaScript dead code affects the energy consumption of mobile devices (i.e., when considering the client side only).

TABLE 7: Results of the further analysis – the (*) symbol denotes statistically-significant correlations

| Variable | Optimization level | P-value | Corr. Coeff. |
|---|---|---|---|
| **Number of dead functions and saving in terms of energy consumption** | | | |
| Energy (mJ) | OL-1 | 0.239 | - |
| | OL-2 | 0.318 | - |
| | OL-3 | 0.830 | - |
| **Number of dead functions and saving in terms of performance** | | | |
| Page load time (ms) | OL-1 | 0.475 | - |
| | OL-2 | 0.125 | - |
| | OL-3 | 0.009 (*) | 0.336 (W) |
| First cont. paint (ms) | OL-1 | 0.432 | - |
| | OL-2 | 0.335 | - |
| | OL-3 | 0.775 | - |
| First paint (ms) | OL-1 | 0.187 | - |
| | OL-2 | 0.134 | - |
| | OL-3 | 0.972 | - |
| **Number of dead functions and saving in terms of network usage** | | | |
| HTTP requests | OL-1 | 0.312 | - |
| | OL-2 | 0.808 | - |
| | OL-3 | 0.781 | - |
| Transferred bytes (Kb) | OL-1 | $4.11 \times 10^{-4}$ (*) | 0.456 (M) |
| | OL-2 | $2.94 \times 10^{-6}$ (*) | 0.603 (M) |
| | OL-3 | $8.16 \times 10^{-6}$ (*) | 0.575 (M) |
| **Number of dead functions and saving in terms of resources usage** | | | |
| CPU usage (%) | OL-1 | 0.886 | - |
| | OL-2 | 0.022 (*) | 0.295 (W) |
| | OL-3 | 0.225 | - |
| GPU usage (%) | OL-1 | 0.643 | - |
| | OL-2 | 0.592 | - |
| | OL-3 | 0.116 | - |
| Memory usage (Mb) | OL-1 | 0.803 | - |
| | OL-2 | 0.108 | - |
| | OL-3 | 0.125 | - |

---

7. The overall experience of a person using a mobile web app, a website, or a computer application, especially in terms of how easy or pleasing it is to use.

As for the **number of HTTP requests and CPU, GPU, and memory usage**, we observed mixed outcomes between in-the-lab and in-the-wild web apps—e.g., we found that different Lacuna optimization levels have an impact on the number of HTTP requests when considering in-the-lab web apps, but not when considering in-the-wild ones. This suggests the possible existence of moderating variables that can diminish or hamper the impact of the Lacuna optimization levels on HTTP requests and CPU, GPU, and memory usage. This could be relevant to *researchers*, who could plan and execute further studies on such moderating variables (also by exploiting our replication package [14]).

Finally, by manually inspecting the source code of the subjects, we conjecture that the emerging results for **HTTP requests and transferred bytes** are due to the fact that web developers tend to statically import and declare JavaScript scripts in their web apps. Specifically, the vast majority of imported/declared scripts are still requested by the browser engine, even though they contain less source code and have a smaller transfer size (due to the dead code being eliminated). This observation highlights the importance of the *bundling technique*, where the number of HTTP requests made to the server is reduced by merging multiple JavaScript files; we advise *web developers* to use existing tools for bundling the JavaScript code of their web apps, such as Webpack [80], gulp-bundle [81], and Browserify [82].

## 5.2 Threats to Validity

We discuss the threats to validity according to Cook and Campbell's categorization [83].

### 5.2.1 Construct validity

We mitigated potential construct validity threats by following well-known guidelines for experimentation in empirical software engineering [49], [50], [51], [52], [53] and by defining all details related to the design of the experiment (e.g., the goal, research questions, tools, variables, statistical analysis procedures) before starting its execution.

### 5.2.2 Conclusion validity

Since all the collected data do not follow a normal distribution, we utilized non-parametric tests. Additionally, we perform the Benjamini-Hochberg correction procedure to account for potential Type-I errors. Finally, we provide a publicly available replication package for independent verification of our findings.

### 5.2.3 Internal validity

A possible threat to the internal validity of the experiment comes from the "maturation" of test subjects, leading them to behave differently across different experiment runs. To mitigate this possible threat, the following precautions have been adopted: (i) the measurement for each experiment trial (i.e., a subject-OL pair) has been repeated 20 times; (ii) the order of execution of the experiment runs has been randomized; (iii) the Chrome app has been cleared and reset before each run so to clean its cache, persisted data, and configuration; (iv) the USB charging of the smartphone is disabled during the execution of each run; (v) between each run the smartphone and the laptop remain idle for 2 minutes to take into account for tail energy usage [76]. Another potential threat to the internal validity comes from the usage of a software power profiler rather than a hardware measurement tool, potentially introducing errors in the measurements. However, the accuracy of the Trepn power profiler has been reported to be close to 99% [72]

### 5.2.4 External validity

To ensure that our experimental subjects are representative of real-world web apps, *in-the-lab* subjects have been complemented with *in-the-wild* subjects. The latter have been sampled from the Tranco list, and constitute a sample of popular real-world websites that are heterogeneous from different perspectives (e.g., application domain, functionalities, size). At the same time, the sample of *in-the-lab* subjects constitutes a varied set of JavaScript development frameworks. Another potential threat to the external validity comes from the usage of a single smartphone device, equipped with an older Android version, for the experiment execution, potentially harming the generalizability of obtained results. This was a forced choice, as the Trepn power profiler does not support newer Android versions. Nonetheless, Trepn is widely used in empirical studies on energy-efficient software [69], [70], [71] which provides confidence in the generalizability of its measurements. Finally, the results of our experiment are obtained by integrating into Lacuna two analyzers: Dynamic and TAJS. The experiment results cannot hold if we use other analyzers for call graph constructions. However, this is true for any choice of analyzer/s to be integrated into Lacuna. Since the experiment on the run-time overhead of JavaScript dead code is very expensive—i.e., 4 optimization levels x 30 subjects x 20 repetitions, in total 2,400 experiment runs—, it is unfeasible to test all the 127 configurations of Lacuna (resulting from the combinations of one to seven analyzers). Therefore, we considered just one Lacuna configuration (i.e., the one based on the joint use of Dynamic and TAJS, suggested by the results of the preparatory experiment (see Section 2.2.7). This choice, taken empirically, should allow for maximizing the benefits resulting from the removal of the dead code.

## 6 RELATED WORK

Dead code (also known as unused code [17], unreachable code [20], and lava flow [16]) has been included in several code-smell catalogs [17], [16], [18] since it is claimed to have negative effects on source code comprehensibility and maintainability [20]. Researchers have investigated the claimed effects of the dead code. For example, Romano *et al.* [84] conducted a controlled experiment where part of the participants had to comprehend and then maintain a Java codebase containing dead code, while another part had to do the same in a codebase deprived of dead code. The authors found that dead code hinders source code comprehensibility, while they could not demonstrate the negative effects of dead code on source code maintainability. Later, Romano *et al.* [19] replicated that experiment three times. The results confirm that dead code penalizes source code comprehensibility; also, they found that dead code negatively affects source code maintainability when developers work on unfamiliar source code. The most important difference between our paper and those introduced just before is that we focus here on the detection and elimination of JavaScript dead code from web apps, while these papers mostly focused on detection and the study the effect of this smell on Java desktop apps. Eder *et al.* [8] conducted a case study on the modifications to dead methods in an industrial web app developed in .NET. Differently from us, the authors only considered dynamic information to detect dead code. In particular, Eder *et al.* monitored the execution of methods in a given time frame, and those methods not executed in a given time frame were considered as dead. In their case study, these authors observed that 48% of the modifications to dead methods were unnecessary (e.g., because dead methods were removed later). A similar finding was reported by Cassieri *et al.* [85] in the context of Java desktop apps hosted on GitHub. This study is different from ours because the authors studied the presence of dead code in Java desktop apps and how developers deal with dead code (e.g., modify and remove it in software evolution tasks).

Researchers have also proposed dead code detection techniques to support developers who aim to remove dead code for refactoring reasons. Chen [86] *et al.* proposed a data model for C++ software repositories supporting reachability analysis and dead code detection. Fard and Mesbah [20] presented JSNOSE, a metric-based technique for detecting smells, including dead code, in JavaScript code. JSNOSE marks a code block as dead if the EXEC metric or the RCH one is equal to zero. The EXEC metric relies on dynamic analysis to count the times a given code block is executed, while the RCH metric measures, by leveraging static analysis, the reachability of a given code block. Boomsma *et al.* [7] proposed a dynamic technique for detecting dead code (dead files, in particular) in web apps written in PHP. This technique monitors the execution of a web app in a given time span to determine the usage of PHP files. A file is deemed as dead if it is not used in that period. The authors applied their technique on a subsystem, allowing the developers to remove 2,740 dead files (i.e., about 30% of the subsystem files). Romano *et al.* [87] proposed DUM, a static technique for detecting dead code (dead methods, in particular) in Java desktop apps, which is based on a call-graph representation where nodes correspond to methods while directed edges correspond to *caller-callee* relationships. The authors implemented DUM in an Eclipse plug-in, named DUM-Tool [88]. Romano and Scanniello [89] explored the use of RTA, an algorithm for call graph construction that is known to be fast and well approximate virtual method calls [90], to detect dead code (dead method, in particular) in Java desktop apps. To this end, they developed a tool, DCF, and evaluated its performance against the one of JTombstone, CodePro AnalytiX, and DUM-Tool. The results of this evaluation show that DCF outperforms the other tools in terms of precision and f-measure of the detected dead methods. As for the recall, DCF is comparable to DUM-Tool. Alabwaini *et al.* [91] proposed a model, based on program slicing, for automatically removing dead code. In particular, they applied a program slicing technique to identify the slices of a program—any code involved in a slice was considered alive. The slices were then merged and any code not involved in a slice was discarded because it was considered dead. The research discussed just before approaches dead code from a refactoring perspective, while we are interested to evaluate the run-time overhead of JavaScript dead code in terms of energy consumption, performance, network usage, and resource usage in the context of web apps.

Researchers have also investigated dead code detection by taking an optimization perspective. Sunitha and Kumar [92] proposed a technique that combines copy propagation and dead code elimination by using hash-based value numbering to avoid executing unnecessary code—e.g., instructions that compute values not used in any execution path starting from them. Karer *et al.* [93] conceived a dead code elimination technique for Java apps based on two steps. First, they converted Java source code into an SSA form—in this form, each variable is assigned exactly once statically. Second, they identified DU-chains to find variables with a definition but without any use during program execution. The found variables are then removed since they are considered dead. Kim *et al.* [94] proposed a technique to efficiently remove dead code in SSA forms, hence obtaining faster and lighter Java bytecode. Wang *et al.* [95] conceived a framework for detecting dead code based on the LLVM compiler infrastructure. The framework consists of three steps. It first translates the source code of the program into an LLVM intermediate representation, then a symbolic execution technique is applied to generate test cases. Finally, the framework combines static and dynamic slicing—the program is analyzed dynamically through the generated test cases—to detect dead code (in particular, dead statements). The proposed framework can be applied to programs written in any programming language as long as it is supported by the LLVM compiler infrastructure. The authors showed that, on five C programs, their framework detected, on average, about 94% of dead statements. Differently from these papers, we present here evidence also about how the presence of JavaScript dead code impacts web apps on Android devices in terms of energy efficiency, loading time, number and payload of HTTP requests, CPU, and memory usage. Vázquez *et al.* [33] proposed a technique called UFFRemover, based on dynamic analysis, to aid developers in identifying and then removing dead functions from the dependencies of JavaScript apps. On the other hand, Lacuna supports both static and dynamic analyses and it is also extensible. Vázquez *et al.* first gathered execution traces of the app being analyzed—for this purpose, the app can be run via its tests in the development environment or via user-app interactions in the production environment—so as to identify the functions that do not belong to any execution trace. These functions are then suggested to developers for removal because they are deemed dead. The authors applied their technique to 22 JavaScript apps and found that around 70% of the functions in the dependencies were dead.

In summary, we contribute in this paper to advance the state of the art on JavaScript dead code identification and elimination in several ways. We can summarize our most important contributions as follows: *(i)* we designed and implemented an extensible approach for JavaScript dead code elimination on which third-party analysis techniques can be reused and integrated and *(ii)* we provide evidence about how JavaScript dead code impacts web apps on Android devices in terms of energy efficiency (slight positive impact), loading time (statistically-significant positive impact), number and payload of HTTP requests (statistically-significant positive impact), CPU and memory usage (mixed results).

# 7 CONCLUSIONS AND FUTURE WORK

In this paper, we present Lacuna, an approach for automatically eliminating JavaScript dead code from web apps. By building on Lacuna, we conducted an empirical evaluation of the run-time overhead of JavaScript dead code in terms of energy consumption, performance, network usage, and resource usage in the context of 30 third-party web apps running on a real Android smartphone. The obtained results lead to relevant implications for users, researchers, and web developers.

As future work, we are planning to extend the formalization of the CG so as to distinguish (and treat differently) between edges that are surely navigated at run-time (e.g., those identified via dynamic analysis) and those that are navigated with a certain probability (e.g., those identified by a static analyzer). We will also expand the CG with the notion of JavaScript module to distinguish between internal, imported, and exported functions and treat them differently while building the CG. Finally, we will integrate additional analysis techniques and tools into Lacuna.

## ACKNOWLEDGMENTS

This project has received funding from the European Union's Horizon 2020 research and innovation programme under the Marie Skłodowska-Curie grant agreement No 871342 "uDEVOPS".

We would like to thank Christos Petalotis and Luka Krumpak, both students of the Vrije Universiteit Amsterdam, for their invaluable help in the external evaluation of Lacuna.